\documentclass[prd,aps,tightenlines,superscriptaddress,showpacs,nofootinbib,eqsecnum,amsfonts,amsmath,epsf]{revtex4}
\usepackage{bm,graphicx,epsfig}
 \def\gsim{\mathrel{
 \rlap{\raise 0.511ex \hbox{$>$}}{\lower 0.511ex
 \hbox{$\sim$}}}}
 \def\lsim{\mathrel{
 \rlap{\raise 0.511ex \hbox{$<$}}{\lower 0.511ex
 \hbox{$\sim$}}}}
\begin{document}
\title{
Parameter estimation of inspiralling compact binaries using 3.5 post-Newtonian
gravitational wave phasing: The non-spinning case}
\author{K.G.\ Arun} \email{arun AT rri.res.in} \affiliation{Raman Research
Institute, Bangalore 560 080, India}
\author{Bala R Iyer} \email{bri AT rri.res.in} \affiliation{Raman
Research Institute, Bangalore 560 080, India}
\author{B.S.\ Sathyaprakash}\email{B.Sathyaprakash AT astro.cf.ac.uk}
\affiliation{School of Physics and Astronomy, Cardiff University, 5, The 
Parade, Cardiff, UK, CF24 3YB} 
\author{Pranesh A Sundararajan}\email{pranesh AT gmail.com} 
\affiliation{Birla Institute of Technology and Science, Pilani}
\begin{abstract}
We revisit the problem of parameter estimation of gravitational-wave 
chirp signals from inspiralling non-spinning compact binaries in the 
light of the  recent extension of the post-Newtonian (PN) phasing 
formula to order $(v/c)^7$ beyond the leading Newtonian order. 
We study in detail the implications of higher post-Newtonian orders 
from 1PN up to 3.5PN in steps of 0.5PN ($\sim v/c$), and examine 
their convergence. In both initial and advanced detectors the estimation 
of the chirp mass (${\cal M}$) and symmetric mass ratio ($\eta$) 
improve at higher PN orders but oscillate with every half-a-PN order.  
In initial LIGO, for a $10 M_\odot$-$10M_\odot$ binary 
at a signal-to-noise ratio (SNR) of 10, the improvement in the estimation of 
${\cal M}$ ($\eta$) at 3.5PN relative to 2PN is $\sim 19\%$ ($ 52\%$). 
We compare parameter estimation in different 
detectors and assess their relative performance in two different 
ways: at a {\it fixed SNR,} with the aim of 
understanding how the bandwidth improves parameter estimation, and 
for a {\it fixed source}, to gauge the importance of sensitivity.  
Errors in parameter estimation at a fixed SNR are smaller for VIRGO 
than for both initial and advanced LIGO. This is because of the 
larger bandwidth over which it observes the signals.  
However, for sources at a fixed distance it is advanced LIGO
that achieves the lowest errors owing to its greater sensitivity.
Finally, we compute the amplitude corrections due to the
`frequency-sweep' in the Fourier domain representation of the
waveform within the stationary phase approximation 
and discuss its implication on parameter estimation. 
We find that the amplitude corrections change the 
errors in ${\cal M}$ and ${\eta}$ by less than 10\% 
for initial LIGO at a signal-to-noise ratio of 10. Our 
analysis makes explicit the significance of higher PN order 
modelling of the inspiralling compact binary on parameter estimation.
\end{abstract}
\date{\today}
\maketitle
\widetext
\section{Introduction}\label{sec:intro}
With  the advent of a new generation of gravitational wave (GW) detectors such as 
LIGO, VIRGO, GEO and TAMA \cite{a1}, we are on the eve of a new era in
astronomy: Gravitational Wave Astronomy (see Ref.~\cite{Cutler-ThorneGR16,GWA} 
for recent reviews). The paucity of GW sources within a
detectable distance, as well as the weakness of the gravitational wave
signals, make imperative the necessity for developing optimal data
analysis techniques, both for their detection and  for the extraction of
maximum information from these signals. It is for this reason that
inspiralling compact binaries, which can be  well modelled within
the general relativistic framework, have become one of the most promising 
candidate sources for the large and medium scale gravitational wave detectors.

An efficient data analysis scheme involves two independent aspects: 
first, the theoretical computation of very high accuracy templates 
and second, the design of a detection strategy adapted to the 
particular signal one is looking for. These strategies vary according 
to the type of signal. 
Gravitational waves from inspiralling binaries are transients lasting for a
short duration in the sensitivity bandwidth of a ground-based detector.
As the binary evolves the waveform sweeps up in frequency and amplitude,
leading to a characteristic {\it chirp} signal.  As the phasing of 
the waves is known very accurately, it is possible to
enhance their detectability by using matched filtering.
Bursts of unknown shape, as for example from a supernova, will be probed 
by monitoring the power excesses in the Fourier or time-frequency domain,
but the enhancement in the visibility of the signal is not as good as when
the phasing of the signal is known and matched filtering can be applied. 
In both cases, coincident observations 
with a network of detectors would assist the detection significantly,
by increasing the confidence level of detection and mitigating non-stationarity.
Continuous sinusoidal signals, as for example from a spinning neutron star,
are also detected by matched filtering and the signal visibility increases
as the square-root of the period for which the signal is observed. 
Stochastic signals require cross-correlation of data from two or more 
collocated, or geographically close by, detectors. Here, the stochastic signal 
buried in one of the instruments acts as a matched filter to dig out {\it exactly}
(or nearly exactly) the same  signal in another. However, since the filter is noisy 
the efficiency is greatly degraded and the visibility improves 
only as the fourth-root of the duration of observation.

As a binary inspirals adiabatically, {\it i.e.}~when the inspiral 
time-scale is much larger than the orbital time-scale, 
it is possible to treat the problem perturbatively and expand the 
general relativistic equations of motion and wave generation as a power 
series in $v/c$, where $v$ is the characteristic orbital velocity of 
the system.  This post-Newtonian (PN) treatment has been successful  
in modelling the dynamics of a binary even at the late stages 
of inspiral and used in the computation of waveforms 
necessary for data analysis (see \cite{Luc-LivRev} for a recent 
review)\footnote{In our nomenclature, $(v/c)^n$ corresponds to 
$\frac{n}{2}$ post-Newtonian (PN) order.  Henceforth, we shall
use units in which $c=G=1.$}.  Since radiation back reaction 
causes the orbital eccentricity $e$ to fall-off, 
for small $e,$ as $e\propto P^{19/18}$ and the orbital radius 
to decay much more slowly $r\propto P^{2/3}$ \cite{peters}, the 
binary  orbit will essentially be circular by the time
the system reaches the late stages of the inspiral phase.
Thus, in our analysis we shall restrict our attention to the
case of compact binaries in quasi-circular orbit, {\it i.e.}~circular 
but for the adiabatic decay of the orbit
under gravitational radiation reaction.
\subsection{Data analysis of the chirp signal: Matched filtering}
Among the different methods suggested for the detection of chirps
from inspiralling and merging binaries, matched filtering 
(also known as Weiner filtering) is the most effective technique \cite{Thorne,Helstrom,Schutz}. 
Matched filtering consists of passing the detector data through a linear 
filter, or a template, constructed from the expected signal $h (t;\bm{\theta}).$ 
Here $\bm{\theta}$ is a `vector' whose components are 
the parameters of the template. The templates $h(t;\bm{\theta})$ generally 
use the restricted waveform where for binaries in quasi-circular orbits the phase
is computed at the highest PN order available, but the amplitude 
is taken to be {\it Newtonian}, involving only the 
dominant signal harmonic at twice the
orbital frequency. This is
different from the complete waveform, which incorporates the PN
corrections to the amplitude, arising from the `plus' and `cross' GW
polarizations, and hence includes the contribution from other 
harmonics (both higher and lower) besides the dominant one. Till date, 
for non-spinning binaries, the restricted waveform is computed to 
3.5PN accuracy \cite{phasing,BFIJ} and the complete waveform up
to 2.5PN order \cite{BIWW,ABIQ1}.  The best template is probably 
the one which consists of the phasing at 3.5PN {\it and} the 
amplitude at 2.5PN.  Presently, both the detection and parameter 
estimation problems mainly employ the restricted PN
waveform although there have been some investigations on
the ensuing improvement achieved when corrections arising from
the other harmonics are incorporated by using the complete waveform 
\cite{Sintes,HM1,HM2}. In this paper, we confine ourselves mostly
to the restricted waveform;  specific amplitude corrections 
arising from  the `frequency-sweep' are considered, however, in  Sec.~\ref{sec:non-RWF}. 

In matched filtering, the unknown set of parameters characterizing
the signal are measured by  maximising the correlation of the 
data with a whole family of templates which correspond to 
different values of the parameters.  The parameters of the template
which maximises the output of a matched filter is an
{\it estimate} of the true parameters. The parameters of a 
signal measured in a single experiment 
will be different from the actual values due to the presence of noise.
Parameter estimation basically aims at computing the 
probability distribution for the measured values of a 
signal. Given a measured value from a single experiment
one then uses the probability distribution function 
to compute the interval in which the true parameters
of the signal lie at a specified confidence level 
(see Sec.~\ref{sec:Overview} for a summary of the theory of parameter 
estimation). In the next Section, we discuss the types 
of error bounds proposed in the literature in the context of GW data 
analysis.
\subsection{Parameter estimation of chirp signal: Different kinds of
error bounds}
In parameter estimation  it is of interest  to
obtain the distribution of the measured values and
error bounds on the measured values of the
parameters. To this end, the starting point would be to construct the
{\it Fisher information matrix}, the inverse of which, the covariance
matrix, provides an estimate of  the possible errors in the measurement
of the parameters \cite{Helstrom}. Error bounds obtained using the
covariance matrix are called the Cramer-Rao bounds  \cite{CRB}.
However, for low values of the signal-to-noise ratio (SNR) 
the actual errors involved may be much 
larger than the errors estimated by this method.  Cramer-Rao 
bounds fall off as the inverse of SNR, whereas the 
actual errors need not follow this behaviour. One usefulness 
of the Cramer-Rao bound is that, they are asymptotically 
valid in the limit of high SNR and hence  provides a 
basis to test all other estimates.

An alternate, and more general, way is to estimate the errors by Monte
Carlo methods \cite{KKT94,BSS1,BSS2}. In this method, one mimics the detection
problem on a computer by performing a large number of simulations
corresponding to different realizations of the noise in each one of
them. The advantage here is that, one  no longer assumes a high SNR, which
is a crucial assumption in computing the covariance matrix. In Ref.
\cite{BSS1} exhaustive Monte Carlo simulations were carried out to
compute the errors in the  estimation of  the parameters and the covariances
among them. It used the initial LIGO configuration  and took into
account only the 1PN corrections assuming, as usual, the orbit to be
quasi-circular. It was shown that the covariance matrix grossly underestimates
the errors in the  estimation of the parameters by over a factor of two
 at a SNR of 10. This discrepancy disappears when
the SNR is approximately 15 for a Newtonian filter and 25
for the 1PN case. Further, the reason for the discrepancy was explained
in detail in Ref. \cite{BaDh98}. Extending the Monte Carlo simulations of
Ref. \cite{BSS1} by  the inclusion of higher order terms would be computationally
quite expensive~\cite{BaDh98}. 

More rigorous bounds (Weiss-Weinstein bound  and Ziv-Zakai bound)
on the parameter estimation of inspiralling binaries are discussed in
Ref. \cite{Nich-Vech}. They compare, at the Newtonian order, the results
obtained by these bounds with the Cramer-Rao bounds and the numerical 
Monte Carlo results. At large SNR, they find all theoretical bounds 
to be identical and attained by Monte Carlo methods. At SNRs below 10, 
the Weiss-Weinstein bound and the Ziv-Zakai bound  provide increasingly 
tighter lower bounds than the  Cramer-Rao bound.
\subsection{Parameter estimation  and the  phasing formula: An update}
Intrinsic parameters, like masses and spins,  characterising the signal 
can be estimated from the data collected by a single detector. On 
the other hand, the distance to the source and its position in the 
sky require at least three geographically separated detectors forming a 
detector network \cite{CF,JK94,JKKT96}. Cutler and Flanagan \cite{CF} have shown that, to a
good approximation, it is sufficient to use Newtonian waveforms in these
analyses. We will not, however, concern ourselves with the estimation of distance
in the present work.

Cutler and Flanagan \cite{CF}  initiated the study of the
implications of higher order phasing formula as applied to the parameter
estimation of inspiralling binaries. They used the 1.5PN phasing
formula to investigate the problem of parameter estimation,
both for spinning and non-spinning binaries, and examined the effect of
the spin-orbit parameter $\beta$ (assumed constant) on the estimation
of parameters. They find that parameter estimation worsens 
by a factor of about ten because of the inclusion of $\beta$. 
The effect of the 2PN phasing formula was analysed
independently by Poisson and Will \cite{PW} and Kr\'olak,
Kokkotas and Sch\"afer \cite{Krolak2}. In both of these works the focus was to
understand the new spin-spin coupling term $\sigma$ appearing at the second
PN order when the spins were aligned perpendicular to the orbital
plane (constant $\beta$ and $\sigma$). Compared to Ref.~\cite{Krolak2}, 
Ref.~\cite{PW} also included the {\it a priori} information about 
the magnitude of the spin parameters, which then leads to a reduction 
in the rms errors in the estimation of mass parameters. It was shown 
that the effect of the  inclusion of $\sigma$ is less drastic than $\beta$
and that it worsens parameter estimation only by a factor of order unity.
In a more recent work \cite{BBW04}, the implications of including the
spin couplings on the parameter estimation and the tests of alternative
theories of gravity were studied using the LISA noise curve.
    
\subsection{Summary of the current work}
Starting with a brief summary of parameter estimation in 
Sec.~\ref{sec:Overview}, we discuss in Sec.~\ref{sec:FourierTransform} 
the nature of the `chirp' signals from {\it non-spinning} 
binaries using the 3.5PN phasing formula \cite{BFIJ}  
which is now completely determined following the recent 
computation of the hitherto unknown parameters at 3PN
\cite{DJS01,BDE04,BDEI04,BI04,BDI04,BDEI05, Itoh}. 

We study parameter estimation using three different noise curves:
advanced LIGO, initial LIGO and VIRGO. Our choice is motivated by the fact
that initial LIGO and VIRGO are the more sensitive instruments among 
the first generation of interferometric detectors with a somewhat different 
combination of bandwidth and sensitivity while advanced LIGO is prototypical
of second generation instruments currently being planned. We will use
the planned design sensitivity curves of initial LIGO and VIRGO as 
in Ref.~\cite{dis3} and advanced LIGO\footnote{For the sake of comparison 
with previous work we have also carried out our study with the 
advanced LIGO noise curve as in Refs.~\cite{CF,PW}. However, most of the 
work reported in this study uses the advanced LIGO noise curve quoted in 
Ref.~\cite{Cutler-ThorneGR16}.} as in Ref.~\cite{Cutler-ThorneGR16}
and discuss in Sec.~\ref{sec:SensitivityCurves} the sensitivity and span of these
instruments for binary coalescences.

As mentioned earlier, Poisson and Will  \cite{PW}  analysed the implications 
of the 2PN phasing formula on parameter estimation of {\it spinning} binaries \cite{KWW}. 
However, extending this to higher orders is not possible at present 
since spin effects beyond 2PN have not yet been computed.  Therefore, in this work we 
will follow the procedure adopted in \cite{PW}, but consider only the  
{\it non-spinning} case. We study in Sec.~\ref{sec:FixedSNR} the effect of 
higher order phasing terms by incorporating them in steps of half-a-PN order
from 1PN up to 3.5PN and examine the convergence of parameter estimation
with PN orders.  We compare the errors for the different noise curves 
and assess their relative performance in two different ways: at a 
{\it fixed signal-to-noise ratio} (Sec.~\ref{sec:FixedSNR}), with the aim 
of understanding how the sensitivity bandwidth improves parameter estimation, 
and for a {\it fixed source} (Sec.~\ref{sec:FixedDistance}), to gauge the 
relative importance of sensitivity and bandwidth.  We have  examined the
correlation of parameter estimation results to the {\it number of useful 
cycles} \cite{dis2} and the sensitivity bandwidth (Sec.~\ref{sec:NUseful}), which together can 
explain the performance of different detectors with regard to parameter estimation.

In Sec.~\ref{sec:non-RWF} we study the effect of the amplitude terms arising from the
`frequency-sweep' $dF/dt$ within the stationary phase approximation
\cite{SPA}. These corrections cause the SNR (which is related to the
total energy emitted by the system) of a given binary to vary as we
go from lower to higher PN orders.  The results are compared 
against the standard restricted waveform approach
and should be viewed as a prelude, {\it albeit} inconsistent, to
parameter estimation using the complete waveform.  We conclude in 
Sec.~\ref{sec:conclusions} with a summary of our results, their regime of
validity, limitations and future directions.

Our main conclusion is that the 3.5PN phasing formula 
leads to an improved estimate of the binary parameters.
For instance, in the case of black hole binaries, at a SNR of 10, the 
estimate of {\it chirp mass} (symmetric mass ratio), more specifically $\ln {\cal M}$ ($\ln\eta$),
improves  while using the 3.5PN phasing 
formula as compared to the 2PN by about 19\% (52\%). Improvements are seen in all cases but are
relatively smaller for lighter binaries. 
At a fixed SNR, VIRGO provides a better estimate of 
the parameters compared to both initial and advanced LIGO configurations
owing to its better sensitivity bandwidth. This is true over the entire
mass range and even for lower mass binaries for which VIRGO accumulates 
fewer number of useful cycles.  For a fixed source, however, advanced LIGO
measures the parameters most accurately, as expected,
with VIRGO doing better than initial LIGO.
Our investigation of the amplitude corrections 
from `frequency-sweep' within the stationary phase approximation 
finds that the percentage change induced by this effect in parameter estimation is less than 10\% for
initial LIGO at a SNR of 10.

\section{A brief summary of  parameter estimation theory}\label{sec:Overview} 
A  firm statistical foundation to the theory of gravitational wave 
data analysis was laid down by the works of e.g.~Finn and Chernoff
\cite{Finn,Finn-Chernoff} and Cutler and Flanagan \cite{CF}. This Section
briefly outlines the problem of parameter estimation relevant to this paper.
Notation and treatment of this Section  essentially follow
Ref.~\cite{Luc-Sathya,CF,PW} (see also \cite{Wainstein,Helstrom,Davies,Krolak1} for
further details).  We restrict our discussion to measurements made by 
a single detector.
\subsection{Matched filtering}
\label{sec:MatchedFiltering}

The output of a gravitational wave detector contains both the signal
and noise and is schematically represented as
\begin{equation}
x(t)=h(t)+n(t)\,,
\label{eq:output}
\end{equation}
where $x(t)$ is the signal registered and $n(t)$ is the noise, 
which is assumed to be a stationary Gaussian random variable, with zero 
mean, {\it i.e.}, \begin{equation}
\overline{n(t)}=0.
\end{equation}
Here an overbar denotes the ensemble average (over many realisations
of the noise or, equivalently, over an ensemble of detectors).
Let $q(t)$ define a linear filter and $c(t)$ its 
correlation with the detector output $x(t)$ 
\begin{equation}
c(t)=\int^{\infty}_{-\infty}dt'\,x(t')\,q(t+t')\,.
\label{eq:correlation}
\end{equation}
Define a new quantity $\sigma[q](t)$, such that $c(t)$ is normalized 
by the square root of its variance, 
\begin{equation}
\sigma[q](t)=\frac{c(t)}{\left [\overline{c^2(t)}-\overline{c(t)}^2\right]^{1/2}}
=\frac{2\Re \int^{\infty}_{0}df\,\tilde{x}(f)\,\tilde{q}^{*}(f)\, e^{2\pi ift}}
{\left[\int^{\infty}_{0}df\,S_h(f)\,|\tilde{q}(f)|^2\right]^{1/2}},\label{eq:sigma}
\end{equation}
where $\tilde{x}(f)$ and  $\tilde{q}(f)$ are  the Fourier transforms of $x(t)$ and 
$q(t),$ respectively, $S_h(f)$ is the real, {\it one-sided} power spectral density defined 
only for positive frequencies by
\begin{equation}
\overline{n(f)\tilde{n}^{*}(f')}=\frac{1}{2} \delta(f-f')\,S_h(f)\,,
\end{equation} 
and
$\tilde{n}(f)$ is the Fourier transform of 
$n(t)$ defined as $\tilde n(f) = \int_{-\infty}^\infty dt\, n(t) e^{-2\pi i f t}$.
The filtered SNR is defined by the ensemble average
 \begin{equation}
\rho[q](t)=\overline{\sigma[q](t)}=\frac{2\Re\int^{\infty}_{0}df\,
\tilde{h}(f)\,\tilde{q}^{*}(f)\,e^{2\pi ift}}
{\left[\int^{\infty}_{0}df\,S_h(f)|\tilde{q}(f)|^2\right]^{1/2}}.
\end{equation}
An optimal filter is the one which maximises the SNR at a particular instant, say $t=0,$ 
and is given by the matched filtering theorem as
\begin{equation}
\tilde{q}(f)=\gamma\, {\tilde{h}(f)\over S_h(f)}\,,\label{eq:filter}
\end{equation}
where $\gamma$ is an arbitrary real constant.
Thus, the SNR corresponding to the optimal filter is given by
\begin{equation}
\rho^2=4\int^{\infty}_{0}df\, \frac{|\tilde{h}(f)|^2}{S_h(f)}.\label{eq:SNR}
\end{equation}
\subsection{Parameter estimation}
\label{sec:ParameterEstimation}
Though we may have a prior knowledge of the {\it form} 
of the signal we will not know what its parameters are. 
Indeed, the parameters are to be measured in the process of
matched filtering. This is achieved by maximising the correlation in Eq.~(\ref{eq:sigma})
with a whole family of templates corresponding to different values of 
the signal parameters. The parameters of the filter which maximise 
the correlation are the {\it measured} values attributed 
by the analyst to the signal presumed to be buried in the data. 
These parameters need not agree, in general, with the {\it actual}
parameters of the signal since the measured values depend on a 
particular realization of the detector noise.

For a given incident gravitational wave, different realizations 
of the noise will give rise to somewhat different best-fit parameters. 
However, if the SNR is high enough, the best-fit parameters will have 
a Gaussian distribution centered around the actual values. 

Let $\tilde{\theta}^{a}$ denote the `true values' of the parameters 
and let $\tilde{\theta}^{a}+\Delta\theta^{a}$ 
be the best-fit parameters in the presence of some realization of the noise.
Then for large SNR, errors in the estimation of parameters
$\Delta\theta^{a}$ obey a Gaussian probability distribution of the form
\cite{Finn}
\begin{equation}
p(\Delta\theta^{a})=p^{(0)} e^{-{1\over2}\Gamma_{bc}\Delta\theta^{b}\Delta\theta^{c}},
\label{eq:prob-dist}
\end{equation}
where $p^{(0)}$ is a normalization constant.  In the above expression
$\Gamma_{ab} \equiv (h_{a}\,|\,h_{b})$, where 
$h_{a}\equiv \partial h/\partial \theta^a,$
is the {\it Fisher information matrix} evaluated at the
{\it measured} value $\hat{\theta}$ of the parameters $\theta$.  Here,
$(\,|\,)$ denotes the noise weighted inner product. Given any two functions
$g$ and $h$ their inner product is defined as:
\begin{equation}
(g\,|\,h)\equiv 2 \int^{\infty}_{0}df\,\frac{\tilde g^{*}(f)\,\tilde h(f)+\tilde g(f)\,
\tilde h^{*}(f)}{S_h(f)}.\label{eq:inner product}
\end{equation}
Using the definition of the inner product 
one can re-express $\Gamma_{ab}$ more explicitly as
\begin{equation}
\Gamma_{ab} = 2 \int_0^\infty
\frac{\tilde{h}_{a}^*(f) \tilde{h}_{b}(f) +
\tilde{h}_{a}(f) \tilde{h}_{b}^*(f)}{S_h(f)}\; df.\label{eq:gamma-eqn}
\end{equation}

The variance-covariance matrix, or simply the covariance matrix, 
defined as the inverse of the Fisher information matrix, is given by
\begin{equation}
\Sigma^{ab} \equiv \langle \Delta \theta^a
\Delta \theta^b \rangle = ( {\Gamma}^{-1})^{ab},
\label{sigma_a}
\end{equation}
where $\langle \cdot \rangle$ denotes an average over the 
probability distribution function in Eq.~(\ref{eq:prob-dist}). 
The root-mean-square error $\sigma_a$ in the estimation of the
parameters $\theta^{a}$ is
\begin{equation}
\sigma_a =
\bigl\langle (\Delta \theta^a)^2 \bigr\rangle^{1/2}
= \sqrt{\Sigma^{aa}}\,,
\label{eq:sigma_a}
\end{equation}and the correlation coefficient $c^{ab}$
between parameters $\theta^a$ and $\theta^b$ is defined as
\begin{equation}
c^{ab} = \frac{\langle \Delta \theta^a
\Delta \theta^b \rangle}{\sigma_a \sigma_b} =
\frac{\Sigma^{ab}}{
\sqrt{\Sigma^{aa} \Sigma^{bb}}}.
\label{eq:c_ab}
\end{equation}
(There is no summation over repeated indices in Eqs.~(\ref{eq:sigma_a}) and
(\ref{eq:c_ab}).)
As a consequence of their definition the correlation coefficients 
must lie in the range $[-1,1].$ When the correlation coefficient
between two parameters is close to $1$ (or $-1$), it indicates that
the two parameters are perfectly correlated (respectively, anti-correlated)
(and therefore redundant) and a value close to $0$ indicates that
the two parameters are uncorrelated; covariance close to 1 (or $-1$)
among parameters causes a large dispersion in their measurement.

In our analysis we will apply the method outlined above to
three prototypical  systems  normally considered
in gravitational wave studies related to ground-based detectors. 
These include a binary neutron star system (NS-NS), a neutron star-black
hole system (NS-BH) and a binary black hole system (BH-BH).
Throughout our analysis we shall assume that the mass of a NS 
is $1.4M_{\odot}$ and that of a BH is $10M_{\odot}.$
\section{Parameter estimation using the 3.5PN phasing formula}
\label{sec:PE-nonspinning}
Having outlined the essential results from the theory of parameter estimation,
we proceed to address the question of extracting the parameters from the
chirp signal using the 3.5PN phasing formula. Our 
 computation parallels the one by Poisson and Will \cite{PW} except that we
confine our attention to the case of non-spinning binaries
whereas Ref.~\cite{PW} dealt with spinning binaries.
\subsection{Fourier transform of chirp at 3.5PN order}
\label{sec:FourierTransform}
To compute the Fisher information matrix we would need the Fourier transform
$\tilde{h}(f)$ of the signal $h(t)$. (Note that here and below $f$ is 
the Fourier transform variable which should not be confused with $F$, 
the instantaneous frequency of emitted radiation.)
Following earlier works, we employ the stationary phase approximation 
(SPA) to evaluate the Fourier amplitude of the waveform.
Given a function $B(t)= 2\, A(t)\,\cos {\phi(t)}$, 
where $d\,\ln A/dt\,\ll\,d\phi(t)/dt$ 
and $|d^2\phi/dt^2|\,\ll(d\phi/dt)^2$,
the SPA provides the following estimate
of the Fourier transform $\tilde{B}(f)$:
\begin{subequations}
\begin{eqnarray}
\tilde{B}(f)&\simeq&{A(t_f)\over \sqrt{\dot{F}(t_f)}}
e^{i\left[\Psi_f(t_f)-{\pi\over4}\right]}\,,\; f\geq 0\,,\\
{\rm where }\;\; \Psi_f(t)&\equiv&2\pi f t -\phi(t)\,,\\
{\rm and }\;\; \frac{d\phi}{dt}&\equiv&2\pi F(t).
\end{eqnarray}
\label{eq:SPA-approx}
\end{subequations}
In this equation $t_f$ is defined as the time at which
$F(t_f)= f$ and $\Psi_f(t_f)$ is  the value of $\Psi_f(t)$ at $t=t_f$.
Starting from the 3.5PN phasing formula in \cite{BFIJ}, the Fourier
transform has been explicitly calculated  in Refs.~\cite{dis3,dis4}.
This Fourier domain waveform, which forms the basis of our
further calculations, is given by
\begin{equation}
\tilde{h}(f) = {\cal A} f^{-7/6} e^{i \psi(f)},
\label{eq:RWF}
\end{equation}
where ${\cal A} \propto {\cal M}^{5/6} Q(\mbox{angles})/D$, and
to 3.5PN order the phase of the Fourier domain waveform is given by
\begin{eqnarray}
\psi(f)&\equiv& \Psi_f(t_f) -{\pi\over 4} \nonumber\\
& = & 2\pi f t_c-\phi_c-\frac{\pi}{4}
+{3\over 128\,\eta\, v^5}\;\sum_{k=0}^{N}\alpha_k\,v^k,
\label{eq:3.5PN-phasing}
\end{eqnarray}
where $v= (\pi M f)^{1/3}$,  $M=m_1+m_2$ is the total mass 
of the binary, $\eta=m_1m_2/M^2$ is the dimensionless mass ratio and $D$ 
the distance to the binary.  
We shall find it useful in our study
to deal with the {\it chirp mass} defined by ${\cal{M}}=\eta^{3/5} M$
rather than the total mass $M.$ 
The coefficients $\alpha_{k}$'s, $k=0,\ldots,N,$ (with $N=7$ at 3.5PN order)
in the Fourier phase are given by
\begin{subequations}
\begin{eqnarray}
\alpha_0&=&1,\\
\alpha_1&=&0,\\
\alpha_2&=&\frac{20}{9}\,\left( \frac{743}{336} + \frac{11}{4}\eta
\right),\label{eq:alpha2}\\
\alpha_{3}&=& -16\pi,\label{eq:alpha3}\\
\alpha_4&=&10\,\left( \frac{3058673}{1016064} + \frac{5429\,
}{1008}\,\eta + \frac{617}{144}\,\eta^2 \right),\\
\alpha_5&=&\pi\left(\frac{38645 }{756}+ \frac{38645 }{252}\,
\log \left(\frac{v}{v_{l\rm so}}\right) - {65\over9}\eta\left[1  + 
3\log \left(\frac{v}{v_{\rm lso}}\right)\right]\right),\\
\alpha_{6}&=&\left(\frac{11583231236531}{4694215680} - \frac{640\,{\pi
}^2}{3} - 
\frac{6848\,\gamma }{21}\right)
+\eta \,\left( - \frac{15335597827}{3048192} + \frac{2255\,{\pi }^2}{12} 
- \frac{1760\,\theta }{3} +\frac{12320\,\lambda }{9} \right)\nonumber\\
&+&{76055\over 1728}\eta^2-{127825\over 1296}\eta^3-{6848\over 21}
\log\left(4\;{v}\right),\\
\alpha_7 &=&\pi\left(\frac{77096675 }{254016} + \frac{378515
}{1512}\,\eta - \frac{74045}{756}\,\eta^2\right).
\end{eqnarray}\label{eq:alphas}
\end{subequations}
Among the coefficients above, $\alpha_5$ can be 
simplified further. This interesting possibility arises
because  of the cancellation of $v^5$ of the 2.5PN term with that of the overall factor in the denominator of  Eq.~(\ref{eq:3.5PN-phasing}). 
Consequently, all but the $\ln v$ terms in $\alpha_5$ are constants
and can be absorbed in a redefinition of the phase\footnote{
We thank Luc Blanchet for pointing this out to us.}.
Indeed, we find that all our estimations, {\it except} $\Delta\phi_c,$ 
remain unchanged irrespective of whether
we choose $\alpha_5$ as above, or a simplified
one retaining only the $\ln v$ term. 

In the 3PN phasing, until recently there were two undetermined parameters,
$\lambda$ and $\theta$, arising from the
incompleteness of the Hadamard self-field regularisation at 3PN\footnote{The ambiguity parameter $\theta$ occurring at 3PN should not be confused
with the  set of parameters $\theta^{a}$ describing the GW.}.
By dimensional regularisation $\lambda$ and $\theta$ have been now
determined in Ref.~\cite{DJS01,BDE04} and \cite{BDEI04,BI04,BDI04,BDEI05}
respectively, completing the general relativistic compact 
inspiral phasing to 3.5PN order: $\lambda=-\frac{1987}{3080}\simeq
-0.6451$ and 
$\theta=-\frac{11831}{9240}\simeq -1.28$. $\lambda$ has also been
determined by alternative approach \cite{Itoh}.

Following earlier works, we choose the set of independent parameters 
$\bm{\theta}$ describing the GW signal to be 
\begin{equation}
\bm{\theta} = (\ln{\cal A}\,, f_0 t_c\,, \phi_c\,, \ln {\cal M}\,,
\ln \eta),
\label{3.6}
\end{equation}
where $t_c$ refers to the coalescence time, $\phi_c$ refers to the phase at coalescence instant, $f_0$ is a scaling frequency related to the
power spectral density (PSD) of the detectors (see next Subsection).
Note that ${\cal A}$ is taken to be one of the independent parameters.
Computing the Fisher information matrix $\Gamma_{ab}$, whose elements
are given by $( h_{a}| h_{b})$ (where $a$ and $b$ are indices which
run over the parameters), is the first step towards our goal.
The upper cut-off in computing the integrals in Eq.~(\ref{eq:SNR})
and (\ref{eq:gamma-eqn}) is taken to be  the GW frequency at
the last stable circular orbit (LSO) given, for a test mass 
in a Schwarzschild spacetime of mass $M,$ to be 
\begin{equation}
F_{\text{upper}}=F_{\text{lso}}= \left(6^{3/2}\pi M\right)^{-1}.
\end{equation}
We take the lower limit in the integrals to be the seismic cut-off frequency 
$f_s$ of the detector. 

\subsection{Sensitivity and span of LIGO and VIRGO}
\label{sec:SensitivityCurves}
\begin{figure}[t]
\centering
\includegraphics[height=3.1in]{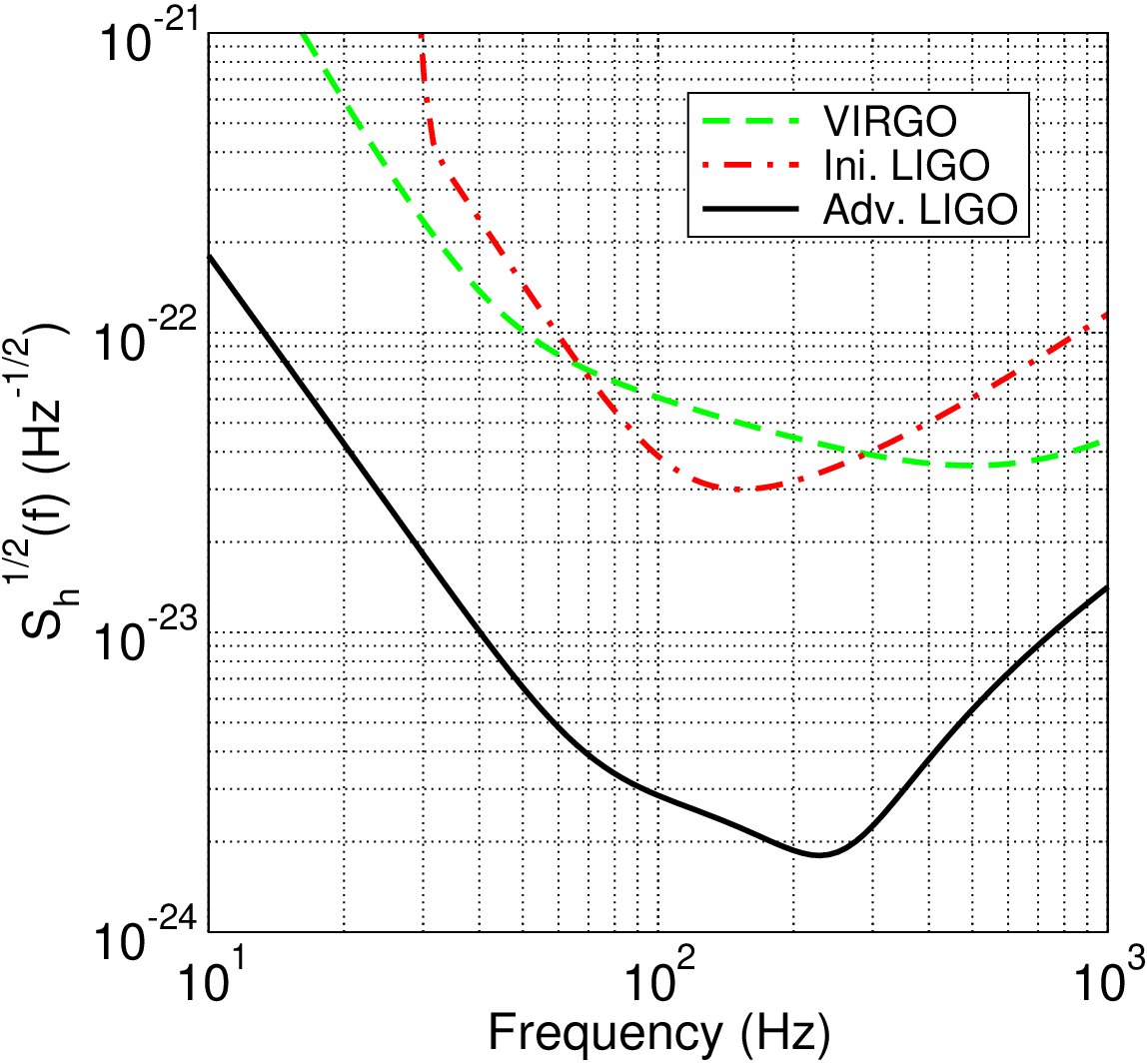}
\includegraphics[height=3.0in]{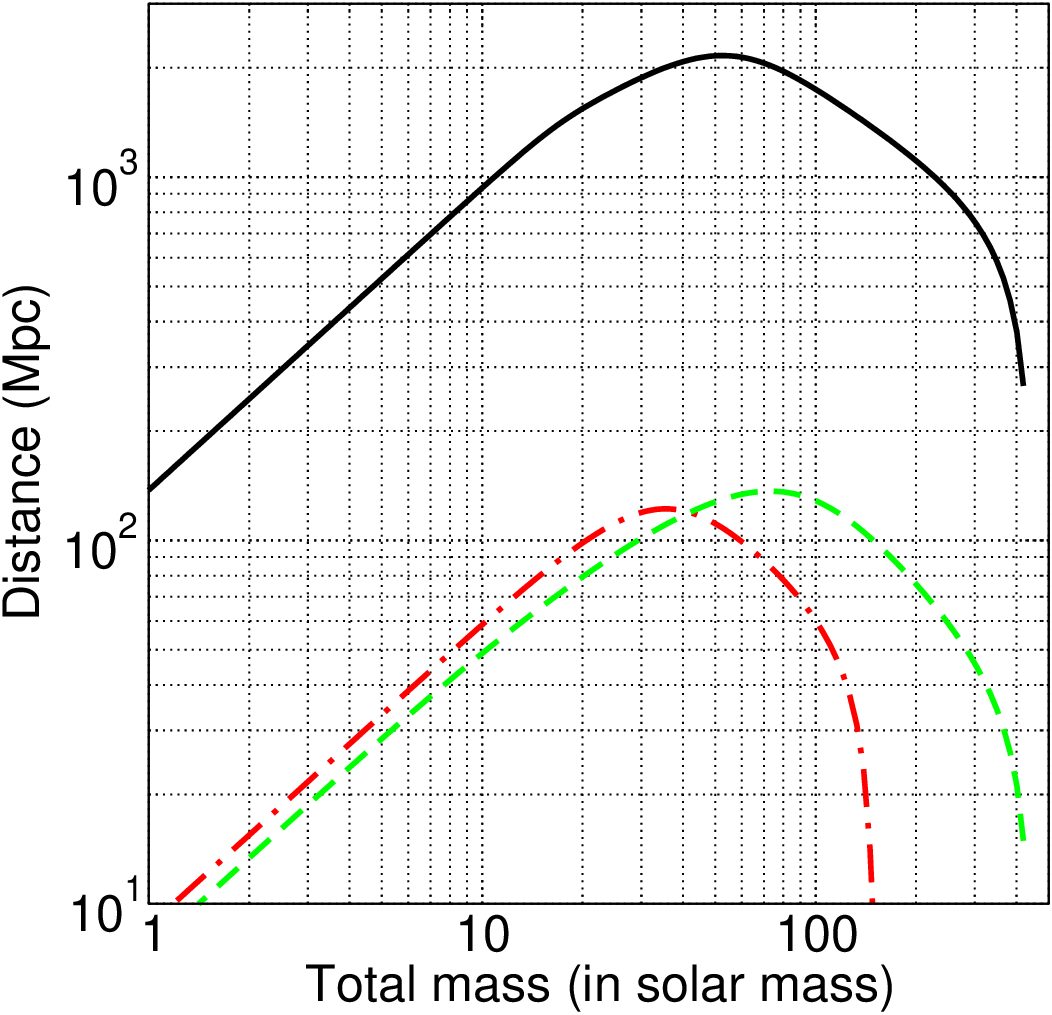}
\caption{Amplitude spectrum (left panel) of initial LIGO, VIRGO and 
advanced LIGO together with the luminosity distance (right panel) at
which RMS-oriented binaries would produce a SNR of $5.$}
\label{fig:NoisePSD}
\end{figure}

We compute the covariance matrix for three noise curves to
understand the effect of detector characteristics on parameter estimation. 
The noise curves used are advanced LIGO as in \cite{Cutler-ThorneGR16}  and
initial LIGO and VIRGO as in \cite{dis3}.  We have fitted the following 
expression to the noise PSD of advanced
LIGO given in \cite{Cutler-ThorneGR16}
\begin{subequations}
\label{eq:AdvLIGOPSD}
\begin{eqnarray}
S_h(f) & = & S_0 \left[ x^{-4.14} - 5x^{-2} + \frac {111 (1-x^2+x^4/2)}{(1+x^2/2)}\right],\; f \geq f_s\\
&=& \infty,\; f < f_s,
\end{eqnarray}
\end{subequations}
where $x=f/f_0,$ $f_0=215$~Hz (a scaling frequency chosen for 
convenience), $f_s=20$~Hz is the lower cutoff frequency [defined
such that for NS-NS binaries the gain in SNR by reducing the lower limit of the
integral in Eq.~(\ref{eq:SNR}) below $f_s$ is less than 1\%], and 
$S_0=10^{-49}\,\text{Hz}^{-1}.$ Note that the above PSD
is significantly different from the advanced LIGO noise curve used
in earlier studies. Indeed, authors of Ref.~\cite{CF,PW,Thorne,Science} use the  
PSD of advanced LIGO to be 
$S_h(f) =  S_0 \left[ x^{-4} + 2 + 2x^2 \right],\;f\geq f_s,$ and $S_h(f)
= \infty,\; f < f_s,$ with $x=f/f_0,$ $f_0=70\,\text{Hz},$ $f_s=10\,\text{Hz}$ 
and $S_0= 6\times10^{-49}\, \text{Hz}^{-1},$ which has a significantly
better low-frequency sensitivity than what is currently believed to be possible
for the next generation of LIGO. Hence, we have chosen to work with the more
recent estimate given in Eq.~(\ref{eq:AdvLIGOPSD}). 

The initial LIGO noise curve from Ref.~\cite{dis3} is given by
\begin{subequations}
\label{ligo-recent}
\begin{eqnarray}
S_h(f) &=& S_0 \left[ (4.49x)^{-56} + 0.16 x^{-4.52} +
0.52 + 0.32 x^2\right],\;f\geq f_s\\
&=& \infty,\; f < f_s,
\end{eqnarray}
\end{subequations}
where again $x=f/f_0$, with $f_0=150$~Hz, $f_s=40$~Hz and $S_0=9 \times 10^{-46}\,\text{Hz}^{-1}$.
Finally, for the VIRGO detector the expected PSD is 
given by \cite{dis3}:
\begin{subequations}
\label{virgo}
\begin{eqnarray}
S_h(f)&=&S_0 \left[ (6.23 x)^{-5} + 2 x^{-1} + 1 + x^ 2 \right],\;f\geq f_s\\
&=& \infty,\; f < f_s,
\end{eqnarray}
\end{subequations}
where $f_0=500$~Hz, $f_s=20$~Hz and $S_0=3.24 \times 10^{-46}\,\text{Hz}^{-1}$.
The amplitude spectra [i.e. the square-root of the power spectral densities 
given in Eqs.~(\ref{eq:AdvLIGOPSD})-(\ref{virgo})] of the various detectors 
are plotted in the left hand panel of Fig.~\ref{fig:NoisePSD}.  

The SNR achieved by these detectors for binaries of different masses not only depends on the
distance $D$ at which the source is located but also on the orientation of the
orbital plane with respect to the line-of-sight. In order not to be biased 
one can consider binaries of root-mean-square (RMS) orientation and compute the
SNR they would produce in a given detector. One can turn around the question 
and ask the distance at which sources of RMS orientation would produce
a specified SNR.  Indeed, the distance $D$ at which a binary of RMS
orientation achieves a SNR $\rho_0$ is given by \cite{dis2}
\begin{equation}
D(M,\eta) = \frac{1}{\rho_0 \pi^{2/3}} \sqrt{\frac{2\,\eta\, M^{5/3}}{15}}
\left [ \int_{f_s}^{f_{\rm lso}(M)} \frac {f^{-7/3}}{S_h(f)}\, df \right ]^{1/2}.
\label{eq:SNRDist}
\end{equation}
As is well known the SNR depends only on the chirp mass ${\cal M}=\eta^{2/3}M$ and not
on the masses of the two bodies separately. The SNR is maximum for equal
mass binaries (for which $\eta=1/4$) and is smaller by a factor $\sqrt{4\eta}$ 
for systems of the same total mass but consisting of stars of unequal masses.
The right-hand panel of Fig.~\ref{fig:NoisePSD} plots the luminosity distance
at which binaries of RMS orientation and consisting 
of stars of equal masses would produce a SNR of $\rho_0=5.$ After computing the
covariance matrix we shall use this plot to study how parameter estimation
varies in different interferometers for sources at a fixed distance.
\subsection{Parameter estimation using 3.5PN phasing -- Fixed SNR}
\label{sec:FixedSNR}
\begin{table}  
\caption{Convergence of measurement
errors  from 1PN to 3.5PN at a SNR of 10 for the three prototypical
binary systems: NS-NS, NS-BH and BH-BH using the phasing formula, in
steps of 0.5PN.
For each of the three detector noise curves the table presents $\Delta
t_c$ (in msec), $\Delta\phi_c$ (in radians), $\Delta
{\cal M}/ {\cal M}$ and $\Delta\eta /\eta$.}
\begin{tabular}{lccccccccccccccccccc} 
\hline 
\hline 
&\vline&
\multicolumn{4}{c}{NS-NS}&\vline&
\multicolumn{4}{c}{NS-BH}&\vline&
\multicolumn{4}{c}{BH-BH}&\\
\cline{3-17}
PN Order&\vline&
$\Delta t_c $&  
$\Delta \phi_c$ &  
$\Delta{\cal M}/{\cal M}$ & 
$\Delta{\cal \eta}/{\cal \eta}$&\vline& 
$\Delta t_c $&  
$\Delta \phi_c$ &  
$\Delta{\cal M}/{\cal M}$&  
$\Delta{\cal \eta}/{\cal \eta}$&\vline& 
$\Delta t_c $&  
$\Delta \phi_c$ &  
$\Delta{\cal M}/{\cal M}$&  
$\Delta{\cal \eta}/{\cal \eta}$
\\ \hline
Advanced LIGO \\\hline
1PN &\vline& 0.3977 & 0.9256 &  0.0267\% &  4.656\%&\vline& 
             0.5959 &  1.261 &  0.1420\% &  7.059\% &\vline& 
             1.162 &  1.974 &  1.041\% &  59.88\% &  \\ 
1.5PN &\vline& 0.4668 &  1.474 &  0.0142\% &  1.638\% &\vline&
              0.7394& 2.091& 0.0763\%& 2.316\%&\vline&
               1.441 &  3.188 &  0.6115\% &  9.609\% & \\ 
2PN &\vline& 0.4623 &  1.392 &  0.0143\% &  1.764\% &\vline&
             0.7208 &  1.848 &  0.0773\% &  2.669\% &\vline& 
             1.404&  2.850 &  0.6240\% &  10.79\% & \\ 
2.5PN &\vline& 0.5090 &  1.359 &  0.0134\% & 1.334\% &\vline&
               0.9000 &  1.219 &  0.0686\% &  1.515\% &\vline& 
               1.819 &  1.574 &  0.5300\% &  5.934\% & \\ 
3PN &\vline& 0.4938 &  1.331 &  0.0135\% &  1.348\% &\vline& 
             0.8087 &  1.131 &  0.0698\% &  1.571\% &\vline& 
             1.544 &  1.580 &  0.5466\% &  6.347\% & \\ 
3.5PN &\vline& 0.5193 &  1.279 &  0.0133\% &  1.319\% &\vline&
             0.9966 &  0.9268 &  0.0679\% &  1.457\% &\vline& 
            2.078 &  1.161 &  0.5241\% &  5.739\% & \\ \hline
Initial LIGO\\\hline
1PN&\vline&  0.3598 &  1.238 &  0.0771\%&  9.792\% &\vline&
             0.9550 &  2.510 &  0.5217\%&  20.06\% &\vline&  
             2.406  &  5.038 &  4.750\% &  216.2\% &  \\ 
1.5PN&\vline&0.4154 &  1.942 &  0.0419\%&  2.768\% &\vline&
             1.182  &  4.135 &  0.2850\%&  5.410\% &\vline&   
             2.986 &  8.143 &  2.781\% &  28.81\% &  \\ 
2PN &\vline& 0.4109 &  1.816 &  0.0423\%  &  3.007\% &\vline& 
             1.148 &  3.597 &  0.2903\%  &  6.316\% &\vline&  
             2.900 &  7.179 &  2.851\% &  32.82\% &  \\
2.5 &\vline&0.4605  &  1.650 &  0.0384\% &  2.129\% &\vline& 
             1.467  &  1.975 &  0.2491\% &  3.305\% &\vline& 
             3.836 &  3.119 &  2.351\% &  16.48\% &  \\
3PN &\vline& 0.4402 &  1.618 &  0.0389\% &  2.170\% &\vline& 
             1.286 &  1.798 &  0.2554\% &  3.474\% &\vline& 
             3.159 &  3.123 &  2.446\% &  17.94\% &  \\
3.5PN &\vline& 0.4754 &  1.517 &  0.0383\% &  2.099\% &\vline& 
             1.666  &  1.324 &  0.2456\% &  3.151\% &\vline& 
             4.512 &  1.912 &  2.314\% &  15.77\% &  \\\hline
 VIRGO\\\hline
1PN &\vline& 0.1363 &  0.5134 &  0.0183\% &  3.044\% &\vline& 
             0.4906  &  1.069 & 0.1134\%  &  5.782\% &\vline& 
             1.621  &  1.854 &  0.8745\% &  52.12\% &  \\
1.5PN &\vline& 0.1578 &  0.7981 &  0.0098\% &  1.004\% &\vline& 
             0.6069  &  1.763 & 0.0603\%  &  1.923\% &\vline& 
             1.430  &  2.972 &  0.5095\% &  8.586\% &  \\
2PN &\vline& 0.1562 &  0.7515 &  0.0098\% &  1.085\% &\vline& 
             0.5918  &  1.561 & 0.0611\%  &  2.215\% &\vline& 
             1.395  &  2.667 &  0.5199\% &  9.625\% &  \\
2.5PN &\vline& 0.1743  &  0.7045 &  0.0091\% &  0.7957\% &\vline& 
             0.7384 &  1.039 &  0.0541\% &  1.263\% &\vline& 
             1.787 &  1.545 &  0.4417\% &  5.370\% &  \\ 
3PN &\vline& 0.1671 & 0.6920 &  0.0092\% &  0.8083\% &\vline& 
             0.6632  & 0.9672 &  0.0551\% &  1.309\% &\vline& 
             1.532 &  1.547 &  0.4552\% & 5.724\% &  \\ 
3.5PN &\vline& 0.1797 &  0.6562 &  0.0091\% &  0.7858\% &\vline& 
             0.8183  &  0.7968 &  0.0536\% &  1.215\% &\vline& 
             2.024  &  1.173 &  0.4369\% &  5.201\% &  \\
\hline\hline\label{table:convergence-nonspinning}
\end{tabular} 
\end{table} 

In this Section, we examine how  the addition of higher order terms in the
phasing formula affects the parameter estimation of the binary. 
We start from the 1PN phasing formula and add terms in steps of half-a-PN
order up to 3.5PN, which is the most accurate expression currently
available. We are interested in the case of non-spinning binaries
(ignoring  spin and orbital angular momentum) and hence estimate only five
parameters $(\ln {\cal A}, f_0 t_c, \phi_c, \ln {\cal M},\ln{\eta})$. 
We calculate the elements of $\Gamma_{ab}$ by explicitly computing the 
derivatives of the Fourier domain waveform  with respect to (w.r.t)
different parameters and 
taking their noise-weighted inner products. The derivatives and the 
Fisher matrices are too lengthy to be displayed here. We note that 
$\Gamma_{1a} = \delta_{1a} \rho^2$, which renders the Fisher information matrix in 
block diagonal form. Since $\ln {\cal A}$ is now entirely uncorrelated with 
all other parameters, we only consider the Fisher  matrix calculated 
from the partial derivatives of $\tilde h(f)$ with respect to the four 
parameters $(f_0 t_c, \phi_c, \ln {\cal M},\ln{\eta})$. 
$\Gamma_{11}$ can be thought of as an independent block, and further 
calculations involving $\cal A$ become trivial.
Finally, by inverting the Fisher information matrix one constructs the 
covariance matrix. 
\begin{figure}[t]
\centering
\includegraphics[width=6in]{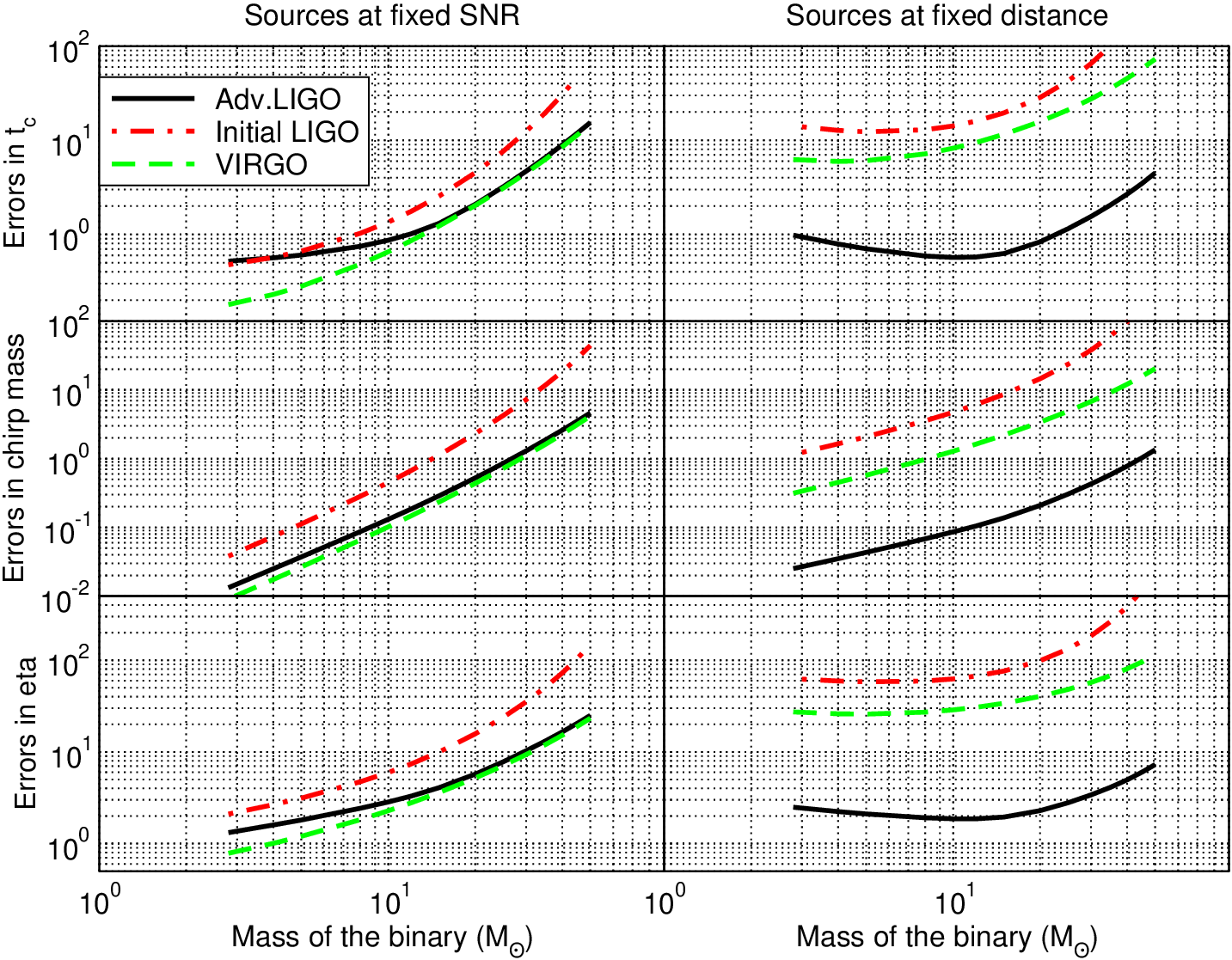}
\caption{Comparison of errors in the estimation of 
$t_c$, ${\cal M}$ and $\eta$ for sources with a fixed SNR 
of 10 (left panels) with those for systems at a fixed 
distance of 300 Mpc (right panels).}
\label{fig:errors-together}
\end{figure}

First, we computed the covariance matrix using the advanced LIGO noise
PSD as defined in Ref.~\cite{CF}, which facilitates a comparison of
our results with those discussed in the literature.
Indeed, at 1.5 PN order we found our results in perfect agreement 
with the numbers given in Table~I of Ref.~\cite{CF} and 
at 2PN order our calculation reproduces the
results in Table~V of Ref.~\cite{Krolak2}.  In both of these papers, 
$\ln \mu$, where $\mu$ is the reduced mass, is chosen to
be the independent parameter instead of $\ln\eta$. However, the errors in
these quantities are simply related by $\frac{\Delta \mu}{\mu}=\frac{2}{
5}\frac{\Delta \eta}{\eta}$, so that the comparison is straightforward.
In the rest of this paper we study only the most recent advanced LIGO
noise PSD together with initial LIGO and VIRGO. 

Next, let us consider the covariance matrix computed using the noise
PSDs of advanced and initial LIGO, and VIRGO, as given in 
Eq.~(\ref{eq:AdvLIGOPSD})-(\ref{virgo}).
The errors in the measurement of the various parameters are tabulated in 
Table~\ref{table:convergence-nonspinning},
for all the interferometers and for three prototypical binaries (NS-NS, NS-BH and BH-BH),
assuming a fixed SNR of 10 in each case. Although the SNR is fixed, different detectors
might accumulate the SNR over different bandwidths, causing the errors to be
greater or smaller compared to one another. In agreement with what one
expects intuitively based on the bandwidth of the various detectors 
(cf.~Fig.~\ref{fig:NoisePSD}, left panel), we find the errors in the various parameters 
to be the smallest for VIRGO, followed by a factor of roughly 10-70\% 
larger errors in advanced LIGO compared to VIRGO, and a factor of 
3 larger errors in initial LIGO compared to advanced LIGO.

In going from lower to higher post-Newtonian order, 
we find that there is an `oscillation' of
the errors in the chirp mass and reduced mass. 
However, the errors at 3.5PN are always smaller than at 2PN.
The opposite oscillation is observed for the errors in $t_c:$
the error in $t_c$ at 3.5PN is always
higher than at 2PN. The fact that the reduced mass and chirp mass
show the same trend is due to the correlation coefficient 
$c_{{\cal M}\eta}$ (listed in Table~\ref{table:correln}) 
all being close to 1.

The oscillation in the variances with PN order can be partially 
understood by an examination of the correlation coefficients 
between $t_c$, ${\cal M}$ and  $\eta$. In Table~\ref{table:correln} we have listed the correlation coefficients together 
with the errors in the estimation of parameters in the case of advanced 
LIGO for a NS-BH system for all PN orders starting from Newtonian but
let us first discuss the trend at orders beyond the 1PN correction.
From this Table we see that 
the estimation of ${\cal M}$ and $\eta$ improves (degrades) 
depending on whether the correlation coefficients $c_{{\cal M}\eta}$
decrease (respectively, increase) with varying PN order. 
Similarly, the estimation of $t_c$ improves (degrades) depending 
on whether the correlation coefficients $c_{t_c{\cal M}}$ 
(or, equivalently, $c_{t_c\eta}$) decrease (respectively, increase)
with PN order. We have also checked that the estimation 
of $\phi_c$  becomes better (worse) with PN order with 
reduction (respectively, enhancement) in the correlation 
coefficients $c_{\phi_c{\cal M}}$ (or $c_{\phi_c{\eta}}$). The same trend is 
seen for other systems and detector configurations, though we do not 
list those numbers to avoid proliferation of details. The behaviour of the errors at 0PN and 1PN is not in agreement with this 
general trend because at 0PN we have only three parameters - $t_c$, $\phi_c$ and 
${\cal M}$. As we go from 1PN to 1.5PN the ambiguity function greatly changes its 
orientation because of the change in sign in the PN series at 1.5PN [cf. 
Eq.~(\ref{eq:alpha2}) and Eq.~(\ref{eq:alpha3})].

Though the PN variation of parameter estimation accuracy 
seems to be dominantly explained by the variation of the 
correlation coefficients, it should be borne in mind 
that the variances in a particular parameter is a combination 
of the covariances and the availability of a greater 
{\it structure} or {\it variety} in the waveforms not
fully assessed in this paper.  This will be the subject 
of a study we shall take up in the near future; it is 
important to understand in more detail why the errors in $t_c$ worsen 
at higher PN orders as it has implications in the
determination of the direction to the source.
\begin{table} 
\caption[]{PN variation in parameter estimation and the associated
correlation coefficients for the NS-BH system for the advanced LIGO
noise
curve.}

\begin{tabular}{lccccccccccccc} 
\hline 
\hline 
PN Order&\vline&$c_{t_c{\cal M}} $&\vline&
$c_{t_c{\eta}}$&\vline& $c_{{\cal M}\eta}$&\vline& $\Delta t_c$ (ms)
&\vline&$\Delta{\cal M}/{\cal M}$ (\%)&\vline&$\Delta \eta/\eta$
(\%)\\\hline
 0PN&\vline&$-$0.6451&\vline&$-$&\vline&$-$&\vline&0.2775&\vline&0.0255&\vline&$-$\\
 1PN&\vline&0.8166&\vline&$-$0.8810&\vline&$-$0.9859&\vline&0.5959&\vline&0.1420&\vline&7.059\\\hline
1.5PN&\vline&0.7983&\vline&0.9280&\vline&0.9444&\vline&0.7394&\vline&0.0763&\vline&2.316\\
2PN
&\vline&0.7947&\vline&0.9239&\vline&0.9460&\vline&0.7208&\vline&0.0773&\vline&2.669\\
2.5PN&\vline&0.8145&\vline&0.9519&\vline&0.9309&\vline&0.9000&\vline&0.0686&\vline&1.515\\
3PN
&\vline&0.8001&\vline&0.9405&\vline&0.9333&\vline&0.8087&\vline&0.0698&\vline&1.571\\
3.5PN&\vline&0.8274&\vline&0.9608&\vline&0.9294&\vline&0.9966&\vline&0.0679&\vline&1.456\\
\hline\hline
\end{tabular}\label{table:correln}
\end{table} 
 Table III summarizes the results of this Section. It provides the
 percentage decrease in the errors due to the greater accuracy
 (3.5PN as opposed to 2PN) in the phasing of the waves: the reduction
 is the highest for a BH-BH binary for which the improvement in
 the estimation of $\eta$ is 52\% and that of $\cal M$ is 19\%
 at an SNR of 10 for the initial LIGO noise curve.

\subsection{Parameter estimation using 3.5PN phasing -- Fixed source}
\label{sec:FixedDistance}
The focus of this Section is to understand the effect of detector 
sensitivity (as opposed to bandwidth) on parameter estimation.  
The results of the previous Section, wherein the errors are quoted
at a fixed SNR, cannot be used to gauge the performance of different 
detectors: a more sensitive detector has a larger SNR for a 
given source and therefore a better estimation of parameters.
Hence, we translate the results for the errors 
in parameter estimation for different detectors but normalized
to a {\it fixed  distance} instead of  a {\it fixed SNR}. 
Since the errors associated with the parameter estimation are inversely
related to SNR ($\sigma\propto1/\rho$), given the error $\sigma_0$
corresponding to a known SNR $\rho_0$ 
(results for $\rho_0=10$ are quoted in Table
\ref{table:convergence-nonspinning}),
one can calculate the error $\sigma$ at another SNR $\rho$ (corresponding to a
fixed distance, say, 300 Mpc) by a simple rescaling of the results
listed earlier. Indeed, $ \sigma=\rho_0\sigma_0/\rho,$ which can be
recast in terms of the distance to the source, using Eq.~(\ref{eq:SNRDist}), as
\begin{equation}\label{eq:sigma-DL}
\sigma(D_L) = \rho_0 \sigma_0 \pi^{2/3} D_L
\left [\frac{2\,\eta\, M^{5/3}}{15} 
\int_{f_s}^{f_{\rm lso}(M)} \frac {f^{-7/3}}{S_h(f)}\, df \right ]^{-1/2}.
\end{equation}

Fig.~\ref{fig:errors-together} summarises the results shown in Table
\ref{table:convergence-nonspinning} (3.5PN entries) over the entire parameter space of interest
for sources with a fixed SNR of 10 (left panels) 
and also the consequent results from the scaling in Eq.~(\ref{eq:sigma-DL})
for  sources at a fixed distance of 300 Mpc (right panels).  
The advantage of having a greater bandwidth
is revealed by looking at panels on the left which shows the errors
in VIRGO to be the smallest, followed by advanced and initial LIGO instruments.
Although the signal-to-noise ratios in the case of VIRGO 
are similar to those of initial LIGO (cf.~Fig.~\ref{fig:NoisePSD},
right panel), Fig.~\ref{fig:errors-together} 
reveals that VIRGO measures the parameters more accurately. 
Indeed, the errors in VIRGO are smaller than in initial LIGO 
by a factor of 2 to 4 and this is entirely as a result of 
VIRGO's larger bandwidth.
Unlike the case of fixed SNR, detector performance
is explicit in the plots for sources at a fixed distance. 
It is evident that the errors reduce by about 30-60 times in advanced
LIGO as compared to initial LIGO. Advanced LIGO gains a factor of 
10-15 in SNR relative to initial LIGO and this accounts for most of 
the improvement in its parameter estimation. However, it also gains
another factor of 3 to 4 because of its greater bandwidth. 
  From the foregoing discussion
we conclude that as far as parameter estimation is concerned
VIRGO performs better than initial LIGO and that advanced LIGO 
can measure the parameters significantly better than what 
one might conclude based on the improvement over VIRGO
in its visibility of the signals. 
\begin{table}
\caption{Percentage change of parameter estimation accuracy
at SNR $\rho=10$  for non-spinning compact binaries due to
improved phasing accuracy from 2PN to 3.5PN. 
Percentage change for the parameter $\sigma_n$ is
taken to be =~$100 \times \left(1 - \sigma_{n}^{\rm
3.5PN}/\sigma_{n}^{\rm 2PN}\right)$. 
Negative values imply worsened parameter estimation in going from
2PN to 3.5PN.}
\begin{tabular}{ccccccccccccccccccccccc} 
\hline 
\hline 
               & \vline & 
\multicolumn{4}{c}{NS-NS}&\vline&
\multicolumn{4}{c}{NS-BH}&\vline&
\multicolumn{4}{c}{BH-BH}\\
\cline{3-16} 
Interferometer & \vline & 
$t_c$ & $\phi_c$ & ${\cal \ln M}$& $\ln\eta$&\vline&
$t_c$ & $\phi_c$ & ${\cal \ln M}$& $\ln\eta$&\vline&
$t_c$ & $\phi_c$ & ${\cal \ln M}$& $\ln\eta$\\ 
\hline 
Adv.~LIGO &\vline&$-$12.33 & 8.118 & 6.993 & 25.23 &\vline & 
$-$38.26 &49.85 &12.16 & 45.41 &\vline&
 $-$48.01 & 59.26 & 16.01 & 46.81 \\
Ini.~LIGO & \vline &
$-$15.70  & 16.47 & 9.456 & 30.20 & \vline&
$-$45.12  & 63.19 & 15.40 & 50.11 & \vline&
$-$55.59  & 73.37 & 18.84 & 51.95\\ 
VIRGO & \vline &
$-$15.05& 12.68& 7.143& 27.58& \vline&
$-$38.27& 48.96& 12.28& 45.15& \vline&
$-$45.09& 56.02& 15.97& 45.96\\
\hline\hline \label{table:improvement}
\end{tabular}  
\end{table}

A final comment:
The plots on the right-hand panel of Fig.~\ref{fig:errors-together} are
somewhat flattened as compared to those on the left-hand panel
due to the fact that errors for sources at a
fixed distance are (anti) correlated with the variation 
of SNR with mass. 
In other words, there are two competing effects on parameter estimation
as the mass of the binary is increased. On the one hand, estimation
becomes worse since the signal spends smaller amount of time in
the detector band and the number of cycles available to discriminate
different signals goes down. On the other hand, as we increase the
mass of the binary the SNR increases thereby aiding in discriminating
between different systems. These competing trends cause the error
in the estimation of the time-of-coalescence and symmetric mass
ratio to show a minimum for a binary of total mass $M\sim 10M_\odot.$
No such minimum is seen, however, in the case of the chirp mass. 
This is because the error in the chirp mass rises more 
steeply with mass than the SNR can cause it to dip.

\subsection{Parameter estimation and Number of useful cycles}
\label{sec:NUseful}
To investigate further the correlation of parameter estimation
performance with detector characteristics we consider the
total number of cycles in the detector bandwidth  and more importantly
the number of {\it useful cycles} for a particular detector
for the three systems under consideration.
The total number of cycles $N_{total}$, is defined as
\begin{equation}
N_{\rm total}=\int^{F_{\rm end}}_{F_{\rm begin}}dF \left({1\over 2\pi}\frac{d\phi}{dF}\right),
\end{equation}
where $\phi$ is the phase of the GW, $F_{\rm begin}$ and $F_{\rm end}$
correspond to the upper and lower cut-off frequencies for the astrophysical
system under consideration. Since the phasing of the waves is a post-Newtonian
expansion in the parameter $v$ the total number of cycles depends on the
post-Newtonian order. At the dominant Newtonian order, assuming that the
lower frequency cutoff of the detector is much smaller compared to the 
last stable orbit frequency of the system, the total number
of cycles for a binary of total mass $M$ and mass ratio $\eta$ is given by
\begin{equation}
N_{\rm total} = \frac{(\pi M f_s)^{-5/3}}{32 \pi \eta}.
\label{eq:NTotal}
\end{equation}
The total number of cycles goes inversely as the mass ratio being the smallest
(for a given total mass) for equal mass binaries and is quite a sharp function
of the total mass. It has an artificiality to it in that  it depends 
on the chosen lower-frequency cutoff, increasing quite rapidly as the 
the cutoff is lowered. Moreover, $N_{\rm total}$ has no information about 
detector characteristics. Motivated by these facts Ref.~\cite{dis2} proposed
that the detector performance can be better understood using the idea of
the {\it number of useful cycles} $N_{\rm useful}$ defined as
\begin{equation}
N_{\rm useful}=\left[\int^{F_{\rm max}}_{F_{\rm min}}\frac{df}{f}\,w(f)\,N(f)\right]\left[ \int^{F_{\rm max}}_{F_{\rm min}}\frac{df}{ f}\,w(f)\right]^{-1}
\label{eq:NUseful}
\end{equation}
where $N(F)$ is the {\it instantaneous number of cycles} (i.e., the number
of cycles spent at the instantaneous frequency $F$) and $w(f)$ is the
weighting function that depends on the effective noise of the 
interferometer and the amplitude of the source defined as
\begin{equation}
N(F)=\frac{F^2}{{dF/ dt}}\,,\ \ \ \ w(f)= \frac{a^2(f)}{h_n^2(f)},
\end{equation}
with $a(f)$ being the `bare amplitude' appearing in the Fourier domain
waveform within the SPA, $|\tilde{h}(f)|\simeq
{a(f)/\sqrt{\dot{F}}}$ and $h_{n}^2\equiv f\, S_h(f).$
Unlike the  total number of cycles, the number of {\it useful cycles} 
contains information about both the detector and the source: it
is  weighted by the noise PSD of the instrument and amplitude
of the source.  Moreover, while the total number of cycles
depends critically on the choice of the lower-cutoff, the number
of useful cycles is a robust estimator and it is pretty much independent
of the cutoffs chosen as long as the frequency range covers the  
sensitivity bandwidth of the instrument.
\begin{figure}[t]
\centering
\includegraphics[width=3.05in]{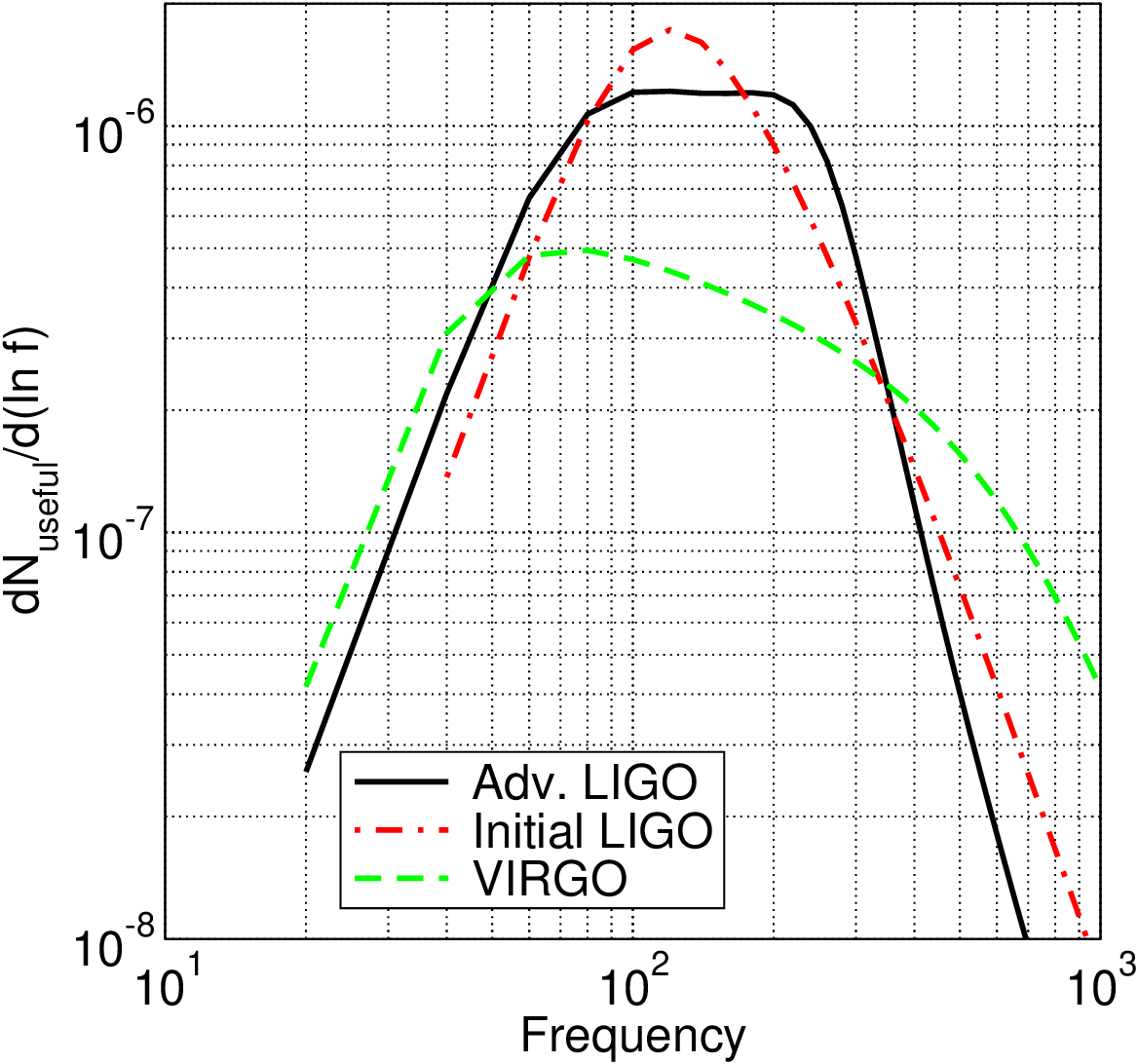}
\hskip 0.1 true cm
\includegraphics[width=3.0in]{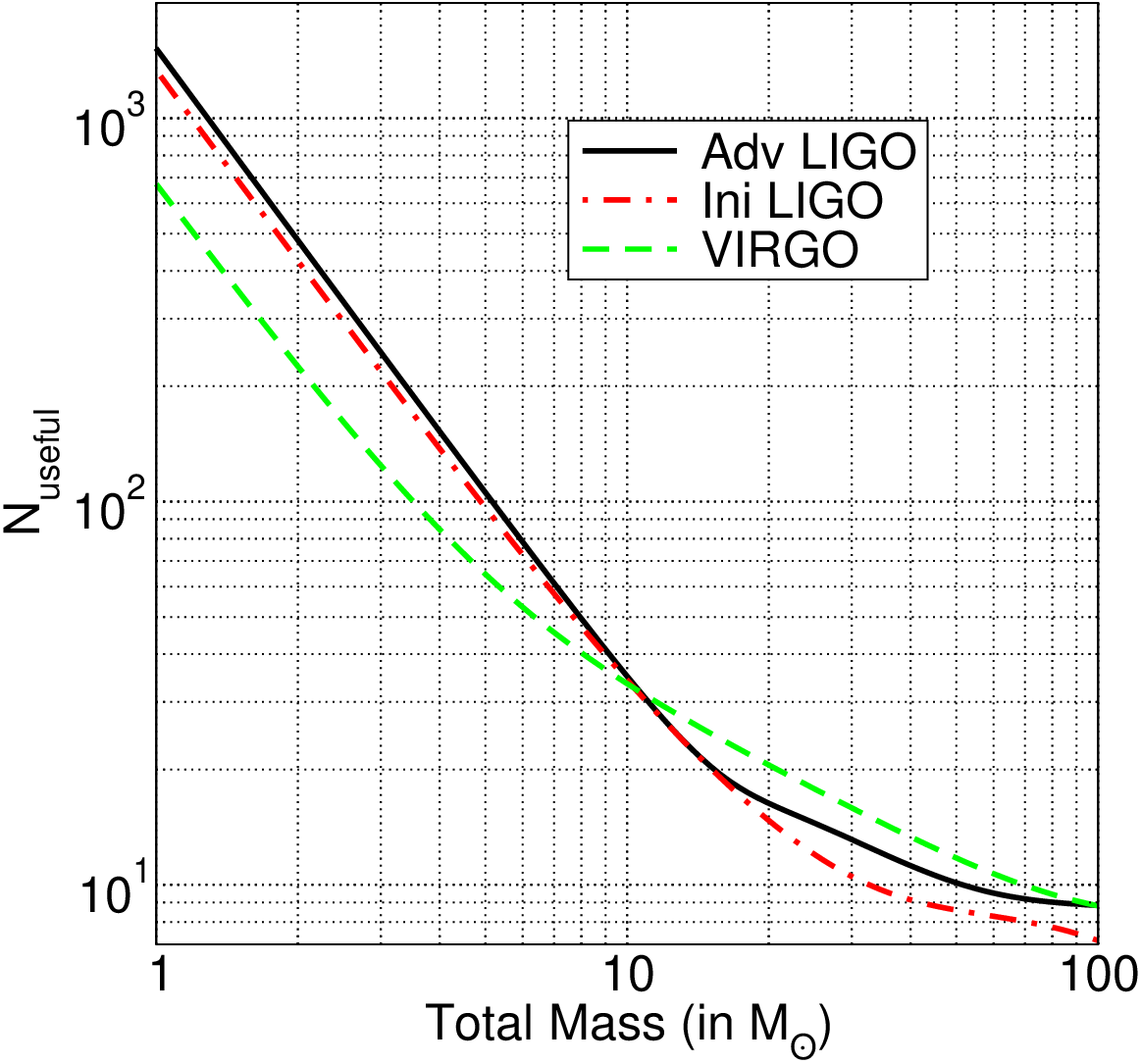}
\caption{Left hand panel is the plot of the  derivative $dN_{\rm useful}/{d(\ln f)}$ against the frequency (in arbitrary normalization) for the three detectors. Similarly, right panel gives the number of {\it useful cycles} as a function of the total
mass of the binary for the three detectors.}
\label{fig:NUseful}
\end{figure}

At Newtonian order, the instantaneous number of cycles is given by
$N(f) = 5 (\pi M f)^{-5/3} / (96 \eta),$ which clearly exhibits the
well-known fact that irrespective of the mass of the system it 
is best to design a detector with a good sensitivity at as 
low a frequency as possible. 
The instantaneous number of cycles decreases rapidly with frequency,
but most of the contribution to the integral in Eq.~(\ref{eq:NUseful})
comes from the region of the band where weighting function 
$w(f) = a^2(f)/h_s^2(f) = f^{1/3}/S_h(f),$ has a minimum. 
As shown in Fig.~\ref{fig:NUseful} (right-hand panel)
for binaries whose total mass is larger than $11\,M_\odot$
the number of useful cycles is larger in VIRGO than the
other two instruments, while just the opposite is true for systems whose
mass is smaller than $11\,M_\odot.$ The reason for this behaviour
can be seen by inspecting the left-hand panel of Fig.~\ref{fig:NUseful}
where we have plotted the integrand $dN_{\rm useful}/d\log f$ of
the number of useful cycles [cf. Eq.~(\ref{eq:NUseful})].
A binary of total mass $100 M_\odot$ has its last stable orbit at
$F_{\text{lso}} \simeq 43 \left (M/100\,M_\odot \right)^{-1}\, \text Hz,$
and increases in inverse proportion to the mass for systems with 
lower masses. Since the integral in Eq.(\ref{eq:NUseful}) is 
terminated at $F_{\rm lso},$ from Fig.~\ref{fig:NUseful} we 
see that as the upper limit of the integral increases (equivalently, the 
mass of the binary decreases) at first the number of useful 
cycles for VIRGO begins to increase. This feature explains why VIRGO
has more number of cycles than the LIGO instruments for binaries
with greater masses. However, owing to their relatively 
narrower bandwidth (as compared to VIRGO) both the LIGO instruments
quickly catch up and for $F_{\rm lso}\gsim300\text Hz$, (equivalently, a total mass of $M \lsim 14\, M_\odot$),
they have greater number of useful cycles than VIRGO.
Thus, the relatively broader bandwidth of VIRGO is responsible for the
smaller number of useful cycles at lower masses.

In general, one can correlate the larger errors associated with the
estimation of parameters of massive systems with the smaller
number of useful cycles  for these systems (see
Table~\ref{table:useful-cycles}).
It may be recalled that Ref.~\cite{dis2} showed that the number of useful
cycles is a good quantifier of detector performance with regard to
detection issues such as effectualness.  However, the efficiency in 
parameter estimation is a combination of bandwidth and the number of 
useful cycles and not the latter alone. Thus, though
VIRGO has a smaller number of useful cycles than the two
LIGO detectors for the NS-NS system, its parameter estimation at a fixed SNR is
far better because of its broader bandwidth. 

Following Ref.~\cite{PW}, where the effects induced in parameter estimation due to
 the inclusion of the 2PN term was understood in terms of the additional total
 number of GW cycles accumulated at that order, we also use a  very similar
 idea to understand the PN variations in parameter estimation of Table~\ref{table:convergence-nonspinning}.
 But unlike \cite{PW}, we use the number of {\it useful cycles} instead of the 
total number of cycles.
 From Table~\ref{table:correlation}, wherein we have given the errors in chirp mass and symmetric mass ratio together with
 the
 contributions to the  useful GW cycles from each PN order term in phasing, it is obvious  that,
 in general,
 when the number of cycles increases in going from one order to another,
 errors decrease (and vice versa) suggesting a possible correlation. Further, following \cite{PW}, we tested this argument by
 artificially
 flipping the sign of each PN order term in the phasing (keeping all lower order terms with the correct sign) and
 comparing the errors. If such a correlation exists, one would expect the trend to be reversed, as the  additional number
 of useful cycles accumulated
 reverses its sign. Indeed Case B of the table
 does show the opposite trend confirming this correlation.
 There is an important exception to this correlation while going from Newtonian to
 1PN, where though  the number of useful cycles increase, the parameter
estimation worsens. A little
 thought reveals that another more dominant aspect comes into play at
 this order due to the inclusion of the new parameter $\eta$ which could
 increase the errors associated with the original set of parameters. This is confirmed by looking at the parameter
estimation of the Newtonian and 1PN orders using a smaller set of four
parameters {\it i.e.} $\{\ln{\cal A}, t_c, \phi_c, \ln{\cal M}\}$, excluding $\ln\eta$. We find that the percentage error in chirp mass decreases from $0.0126$ to $0.0120$ for NS-NS case and $0.4833$ to $0.4183$ in the BH-BH case in step with the
increase in number of useful cycles. 
 However, the reason behind the anomalous behaviour in going from 1 to 1.5PN
 and 3 to 3.5PN -- where despite the decrease in the number of useful cycles,
the parameter estimation improves -- is not clear from the present analysis.
 Thus the previous considerations are not sufficient to completely
 understand the variation of parameter estimation with the PN order.

  Based on the understanding obtained in the previous paragraph, we 
  conclude the Section with the following comment:
  At present we do not have a detailed understanding of the reason 
  underlying the variation in parameter estimation with PN orders since
  the inclusion of higher PN terms could lead to one or more of the following:
  (a)~introduction of a new parameter (e.g.\ $\eta$ in
  going from 0PN to 1PN) leading to an increase in the variance of
  the existing parameters, (b)~increase in the `variety' of waveforms leading
  to a reduction in the variance, and (c)~change in the covariance 
  among the various parameters.  Though by a critical examination 
  of the results summarised in Tables~\ref{table:convergence-nonspinning} 
  and \ref{table:correln} some of these effects can be seen in 
  action, it is not easy to disentangle these individual effects and present
  a consistent quantitative picture.  This, we leave to a future study.

\section{Beyond the restricted waveform: Amplitude corrections due to
frequency-sweep and its implications}\label{sec:non-RWF}
In the foregoing Sections we worked with the restricted PN 
approximation. In this approximation the GW phase is taken to
as high a PN accuracy as available while the amplitude is assumed
to be Newtonian. Indeed, all harmonics, except the dominant one at twice
the orbital frequency, are neglected. From Eq.~(\ref{eq:SPA-approx}), 
one can see that the Fourier-domain amplitude is determined by
the product of the time-domain amplitude $A(F)\propto F^{2/3}$ and
the factor $({dF/dt})^{-1/2},$ where ${dF/dt}$ is the 
`frequency-sweep' or `chirp rate' of the signal. The frequency-sweep 
provides a way of (partially) computing the dependence of the 
wave amplitude on different PN orders.  This correction, in 
addition to being calculable, should be of some relevance when 
we compared in Sec.~\ref{sec:ParameterEstimation} 
parameter estimation accuracy at different PN 
orders where, following Ref.~\cite{PW}, we 
assumed the SNR to be the same at all PN orders. Our 
assumption was justified since in the restricted PN approximation
there is no change in the amplitude of the signal as we go from
one PN order to the next.  However, the
frequency-sweep causes the Fourier amplitude to change across the PN
orders and leads to variations in the SNR with the PN orders. Since
the errors depend on the SNR, one should rescale the errors
by the ratio of SNRs to compare fairly the PN trends in parameter
estimation of the chirp signal. In what follows, 
we will set up the necessary  formulas to normalize the
errors to the same SNR. However, 
it is immediately obvious that a more consistent calculation should
begin with the {\it full} amplitude corrections arising from the GW
polarizations computed in Ref.~\cite{BIWW,ABIQ1}, in lieu of the restricted
 approximation used here, and by including the sub-dominant harmonics. 
Inclusion of these terms is beyond the scope of this paper 
and will be addressed elsewhere.
\begin{table}
\caption{
Number of useful cycles (and total number of cycles in brackets) 
for different systems and different detectors computed using
3.5PN phasing. To compute the total number of cycles the 
lower cut-off is chosen to be the seismic cutoff frequency 
of each detector and the upper cutoff is the frequency 
corresponding to the LSO.}
\begin{tabular}{ccccccc}
\hline
\hline
Detector  &\vline& NS-NS &\vline& NS-BH &\vline& BH-BH \\
\hline
Adv.~LIGO &\vline& 284 (5136)  &\vline&60 (1111)  &\vline&14 (184) \\
Ini.~LIGO &\vline& 251 (1615)  &\vline&59 (330)  &\vline& 12 (52)  \\
VIRGO     &\vline& 140 (5136)  &\vline&64 (1111)  &\vline&18 (184) \\
\hline
\hline
\end{tabular}
\label{table:useful-cycles}
\end{table}
\begin{table}  
\caption{Correlation of parameter estimation and number of useful cycles
 with PN order ($n$) for NS-NS and BH-BH Binaries for initial LIGO noise curve. Case A corresponds to the standard PN coefficients in the phasing formula ($\epsilon_a=1$, $ a\leq n$). Case B  refers to the results corresponding to a flip in sign of the $a=nPN$ term keeping all other lower orders with correct sign ($\epsilon_a=-1$ for $a=n$ and 1 for $ a< n$). Errors listed are all in percentages. The values for the Newtonian order are obtained using a set of four parameters, $\{\ln {\cal A}, t_c,\phi_c, \ln {\cal M}\}$, excluding $\ln\eta$. }
\begin{tabular}{lccccccccccc} 
\hline 
\hline 
&\vline&
\multicolumn{3}{c}{NS-NS}&\vline&
\multicolumn{3}{c}{BH-BH}\\
\cline{3-9}
PN Order ($n$)&\vline&
$\Delta{\cal M}/{\cal M}$ & 
$\Delta{\cal \eta}/{\cal \eta}$&
$N_{useful}$&\vline&
$\Delta{\cal M}/{\cal M} $&  
$\Delta{\cal \eta}/{\cal \eta}$&
$N_{useful}$
\\ \hline
Case A 
\\\hline
0PN&\vline&0.0126 & &247.8&\vline&
0.4833 & & 14.98&\\
1PN &\vline&0.0771  & 9.792&27.13&\vline&
    4.750 & 216.2 &  7.283\\
1.5PN &\vline&0.0419  & 2.768& $-$22.98&\vline&
    2.781 & 28.81&  $-$9.148\\
2PN&\vline&0.0423 & 3.007& $-$1.197 &\vline&
    2.851 & 32.82& $-$0.496\\
2.5PN&\vline&0.0384 & 2.129& 2.406&\vline&
    2.351 & 16.48& 1.850\\
3PN&\vline& 0.0389 & 2.170& $-$1.735&\vline&
  2.446 & 17.94& $-$1.971\\
3.5PN&\vline& 0.0383 & 2.098& $-$0.151&\vline&
   2.313 & 15.75& $-$0.236\\

\hline
Case B
\\\hline
1PN&\vline&0.0771 & 8.858& $-$21.65&\vline&
  4.750 & 158.2& $-$3.579\\
1.5PN&\vline&0.0547 & 1.842& 80.65 &\vline&
  3.237 & 28.04& 21.56\\
2PN &\vline& 0.0415 & 2.564& $-$53.64&\vline&
   2.727 & 25.63& $-$19.30\\
2.5PN &\vline& 0.0515 &5.085& $-$4.700&\vline&
   14.96 & 473.7& $-$2.395\\
3PN &\vline& 0.0380 & 2.089& 6.563&\vline&
  2.271 & 15.26&  6.461\\
3.5PN &\vline&0.0395 & 2.248& $-$3.453&\vline&
  2.625 & 20.89&$-$4.978\\
\hline\hline\label{table:correlation}
\end{tabular} 
\end{table}

To estimate the amplitude corrections due to the frequency-sweep
$\dot{F}$, we
start from the Fourier domain waveform in the  stationary phase approximation
which can be written as
\begin{equation}
\tilde{h}(f) \equiv \int_{-\infty}^\infty h(t)\, e^{-2\pi i f t}\,
dt=\left[2\eta\,{M\over d}\,Q(\text{angles})\right]
\frac{v^2}{\sqrt{\dot{F}(v)}} e^{i\,\psi(f)},
\label{eq:full-SPA-FD}
\end{equation}
where $v=(\pi M f)^{1/3}$.  Using the expression for $\dot F$ 
at the Newtonian order, it can easily be shown that 
Eq.~(\ref{eq:full-SPA-FD}) reduces to Eq.~(\ref{eq:RWF}).  From 
Eq.~(\ref{eq:full-SPA-FD}) it is clear that the PN
corrections in the frequency-sweep $\dot F$ [see Eq.~(\ref{eqn:fdot6a})
below] introduces a related PN correction in the amplitude as discussed
earlier in the Section. To proceed further we note that the 
formula for $\dot F$ can be
normalized w.r.t. its Newtonian value ${\dot F}_{\cal N}$
and written as the product of the Newtonian value and PN corrections
${\dot F}_{\cal C}$:
\begin{equation}\label{eq:f-calf}
\dot{F} = \dot{F}_{\cal N}\;{\dot F_{{\cal C}}}.
\end{equation}
Schematically $\dot F_{{\cal C}}$ can be written as 
\begin{equation}
\dot F_{{\cal C}}=\left[1+\dot F_{{\cal C}}^{1PN}+\dot  F_{{\cal C}}^{1.5PN}+\dot
F_{{\cal C}}^{2PN}+\dot F_{{\cal C}}^{2.5PN}+\dot F_{{\cal C}}^{3PN}+ \dot
F_{{\cal C}}^{3.5PN}+\cdots\right].
\end{equation}
Using $F_{{\cal N}}=\frac{96}{5\pi\,{\cal M}^2}\;(\pi {\cal M}
F)^{11/3}$ and $v=(\pi M F)^{1/3}$ and some simple algebra, one can
write,
\begin{equation}
\tilde{h}_{{\cal C}}(f)= {\cal B}_{\cal N}\,{\cal B}_{{\cal
C}}\,e^{i\psi(f)},\ \ \ \ 
{\cal B}_{\cal N} = {\cal A}\,f^{-7/6},\ \ \ \ 
{\cal B}_{{\cal C}} = {1\over \sqrt{\dot F_{{\cal C}}}}
\label{eq:h-corrected}
\end{equation}
where ${\cal B}_{\cal N},$ as in Eq.~(\ref{eq:SNR}), is the 
Newtonian functional dependence.
Using Eq.~(\ref{eq:h-corrected}), the expression for SNR can be re-written as
\begin{equation}
\rho^2=4\int_{0}^{\infty}df\;{\cal B_{\cal C}}^2\,{ {\cal B}_{{\cal
N}}^2\over S_h(f) }\label{eq:rho-PN}
\end{equation}
From the definition of SNR, Eq.~(\ref{eq:SNR}), it is clear that the  SNR
varies with the PN order of $\dot F_{{\cal C}} $. Similarly, one can
write down the components of the Fisher matrix $\Gamma_{ab}$ as 
\begin{equation}
\Gamma_{ab}=2 \,\int^{+\infty}_{0}df\;\frac{{\cal
B_{\cal N}}^2}{S_h(f)}\;\left[{\partial{\cal B_{C}}\over
\partial{\theta_a}}{\partial{\cal B_{C}}\over \partial{\theta_b}}+{\cal
B_{C}}^2{\partial{\psi}\over \partial{\theta_a}}{\partial{\psi}\over
\partial{\theta_b}}\right]\label{eq:gamma-AC}
\end{equation}
where $\theta_{a}$ and $\theta_{b}$ are the parameters in the GW signal.
(${\cal B}_{{\cal C}}$ is a PN series in $v$ and its $\theta$ dependence
arises solely from the mass dependence of $v$). 
In Sec~\ref{sec:PE-nonspinning}, ${\cal B_{C}}$ was effectively
taken to be unity. Here we relax  that assumption by taking 
into account the PN corrections involved.

The frequency-sweep appearing in Eq.~(\ref{eq:full-SPA-FD}) above can 
be straightforwardly calculated from the expressions for the flux function 
${\cal F}$ and the energy function ${E}$, determining the GW phasing in the
adiabatic approximation. It is given by \cite{dis3}
\begin{equation}
\dot F(v) = -\frac{3v^2}{\pi M^2} \frac{{\cal F}(v)}{E'(v)},
\end{equation}
where $v= (\pi M f)^{1/3}$ and ${ E'=\frac{dE}{dv}}$. Using the 3.5PN
accurate expression for $E$ and ${\cal {F}}$ available in 
\cite{BFIJ}, the expression for $\dot F$ up to 3.5PN is given by
\begin{eqnarray}
\left(\frac{dF}{dt}\right)^{3.5PN} &=& \frac{96}{5\pi {\cal M}^2} (\pi
{\cal M} F)^{11/3}
\biggl[ 1 - \biggl( \frac{743}{336} + \frac{11}{4} \eta \biggr)
(\pi M F)^{2/3} + (4 \pi) (\pi M F)
\nonumber \\ & & \mbox{}
+ \biggl( \frac{34103}{18144} + \frac{13661}{2016}\eta +
\frac{59}{18} \eta^2  \biggr) (\pi M F)^{4/3}+\left(-{4159\pi\over672}
-{189\pi\over 8}\eta\right)(\pi M F)^{5/3}\nonumber\\
&&+\left[{16447322263\over
139708800}+{16\pi^2\over3}-{1712\over105}\gamma+\left(-{273811877\over
1088640}+{451\pi^2\over48}-{88\over3}\theta+{616\over9}\lambda\right)\eta\nonumber\right.\\
&&+\left.{541\over896}\eta^2-{5605\over2592}\eta^3-{856\over105}
\log\left(16\,{x}\right)\right](\pi M F)^{2}\nonumber\\
&&+\left(-{4415\over4032}+{358675\over6048}\eta+{91495\over1512}\eta^2\right)\pi
(\pi M F)^{7/3}\ \biggr].\label{eqn:fdot6a}
\end{eqnarray}
In the above expression $\gamma$ is  the Euler's constant
($\gamma=0.577\cdots$) and the coefficients
$\lambda=-\frac{1987}{3080}\simeq -0.6451$, 
$\theta=-\frac{11831}{9240}\simeq -1.28$.
\begin{table} 
\caption{Variation in SNR with the PN order due to the
amplitude corrections arising from the frequency-sweep 
$dF/dt$ in the stationary phase approximation. We
assume initial LIGO noise spectral density and
place the source such that at 0PN order we have a SNR of 10.}
\begin{tabular}{lcrcrcr} 
\hline 
\hline 
PN Order&\vline & NS-NS &\vline& NS-BH &\vline& BH-BH \\
\hline 
0PN   &\vline&10.00 &\vline& 10.00 &\vline& 10.00 \\
1PN   &\vline&10.53 &\vline& 11.26 &\vline& 12.19 \\
1.5PN &\vline&10.08 &\vline&  9.560&\vline&  9.357 \\
2PN   &\vline&10.06 &\vline&  9.483 &\vline&  9.178 \\
2.5PN &\vline&10.11 &\vline&  9.736 &\vline&  9.861 \\
3PN   &\vline&10.08 &\vline&  9.355 &\vline&  9.157 \\
3.5PN &\vline&10.07 &\vline&  9.341 &\vline&  9.060 \\
\hline 
\hline 
\label{table:percentDC} 
\end{tabular} \label{table:SNR-PN} 
\end{table} 

 In
Table~\ref{table:SNR-PN}, we summarize how the SNR varies with the PN order
for different sources assuming that the SNR corresponding to the Newtonian 
order is $10.$ The  convergence of the SNR's with PN orders is pretty 
obvious, although it should be recalled that the complete waveform
includes PN corrections from other harmonics that are comparable to
the higher order terms in the frequency-sweep \cite{BIWW,ABIQ1}.
It would be interesting to see how the results change when these 
are included.  We also note that the variation of the SNR is 
greater for systems with larger masses. Using a 3.5PN
frequency-sweep, instead of the Newtonian one, increases the 
SNR by $0.7\%$ for a NS-NS binary, while the SNR decreases 
by $9.5\%$ for a BH-BH binary.
Though these amplitude corrections may not be important for NS-NS 
binaries, they might be relevant for the BH-BH case.

Using the results in Table~\ref{table:SNR-PN} one can implement a simple 
procedure to obtain  better error estimates. One can scale the results of 
Sec.~\ref{sec:PE-nonspinning}, obtained within  the restricted waveform
approximation, by the factor $\rho_{n}/\rho_{0}$, where $\rho_{n}$ and 
$\rho_{0}$ are the SNRs at $n$PN and 0PN orders, respectively. In this 
simple estimate one is effectively neglecting the contributions to the 
Fisher matrix from the variation of the $\dot{F}$ terms in the amplitude 
w.r.t the signal parameters $\bm{\theta}$ (see Eq.~(\ref{eq:gamma-AC}). 
We incorporate this contribution in a more general and rigorous way 
in what follows.

Our more general procedure is based on Eqs.~(\ref{eq:rho-PN}) and
(\ref{eq:gamma-AC})  which accounts for the SNR and the Fisher matrix,
respectively, with the full $\dot{ F}$ dependence in amplitude.
The steps leading to  the final results listed
in Table~\ref{table:convergence-nonspinning1-AC} are as follows:
(i) compute the amplitude ${\cal A}$ such that the SNR at 0PN is 10;
(ii) compute the Fisher matrix  taking into account the
amplitude corrections from the  frequency-sweep using 
Eq.~(\ref{eq:gamma-AC}); (iii) scale the final results by 
$\rho_{n}/\rho_{0}.$ The covariance matrix obtained 
from such a procedure can then be compared with that  obtained  in
Sec.~\ref{sec:PE-nonspinning}.  The procedure  above is obviously equivalent 
to choosing a `running' amplitude ${\cal A}_{n},$ with $\rho_{0}=10.$ 

In Table~\ref{table:convergence-nonspinning1-AC}, the variation of errors 
with different PN orders is shown for the initial LIGO noise curve\footnote{The numbers listed in
Tables~\ref{table:SNR-PN} and \ref{table:convergence-nonspinning1-AC}
are those obtained by numerically integrating Eqs~(\ref{eq:rho-PN}) and (\ref{eq:gamma-AC}) {\it without}
 any further re-expansion of ${\cal B}_{C}$ in Eq.~(\ref{eq:h-corrected}).}.
The oscillation of errors with PN orders remains after the
inclusion of the frequency-sweep and one infers that changes due 
to these amplitude terms are not very
significant. At an SNR of 10 the difference is at most 10\%.
\begin{table}[t]
\caption{Parameter estimation with amplitude corrections from
the frequency-sweep incorporated for  the initial LIGO
noise curve and SNR $\rho=10$.  $n$PN refers to
the choice of $n^{th}$ PN order both in the amplitude and phase
of the frequency-domain waveform.}
\begin{tabular}{lcrrrrcrrrrcrrrr}
\hline
\hline
& \vline & \multicolumn{4}{c}{{NS-NS}} 
& \vline & \multicolumn{4}{c}{{NS-BH}} 
& \vline & \multicolumn{4}{c}{{BH-BH}} \\
\cline{2-16} 
PN order & 
\vline & 
$\Delta t_c$ & $\Delta \phi_c$ & $\Delta {\cal M} / {\cal M}$ & $\Delta
\eta / \eta$ & 
\vline &
$\Delta t_c$ & $\Delta \phi_c$ & $\Delta {\cal M} / {\cal M}$ & $\Delta
\eta / \eta$ &
\vline &
$\Delta t_c$ & $\Delta \phi_c$ & $\Delta {\cal M} / {\cal M}$ & $\Delta
\eta / \eta$ \\
\hline
1PN   &
\vline& 0.3208 & 1.157 & 0.0752\% & 9.431\% & \vline
      & 0.9077 & 2.449 & 0.5241\% & 19.94\% & \vline
      & 1.717  & 3.544 &3.400\%   & 152.6\% \\
1.5PN &
\vline& 0.4417 & 2.018 & 0.0426\% & 2.851\% & \vline
      & 1.234  & 4.232 & 0.2840\% & 5.484\% & \vline
      & 2.929  & 7.883 &2.647\%   & 27.71\% \\
2PN   &
\vline& 0.4427 & 1.903 & 0.0432\% & 3.115\% & \vline  
      & 1.206  & 3.695 & 0.2892\% & 6.408\% & \vline
      & 2.830  & 6.897 &2.680\%   & 31.24\% \\  
2.5PN &
\vline&0.4693  & 1.672 & 0.0387\% & 2.149\% & \vline
      &1.496   & 2.009 & 0.2480\% & 3.326\% & \vline
      &3.801   & 3.093 & 2.316\%  & 16.28\% \\
3PN   &
\vline& 0.4768 & 1.709 & 0.0398\% & 2.261\% & \vline
      & 1.387  & 1.919 & 0.2556\% & 3.583\% & \vline
      & 3.106  & 3.068 & 2.307\%  & 17.20\% \\
3.5PN &
\vline& 0.5186 & 1.615 & 0.0393\% & 2.199\% & \vline
      & 1.787  & 1.435 & 0.2459\% & 3.260\% & \vline
      & 4.474  & 1.939 & 2.203\%  & 15.33\% \\
\hline
\hline
\end{tabular}
\label{table:convergence-nonspinning1-AC}
\end{table}
\section{Conclusion}
\label{sec:conclusions}
\subsection{Summary and discussion of results}
We have carried out a detailed  study to understand the
implication of 3.5PN phasing formula on parameter estimation of
non-spinning binaries using  the covariance matrix.
We also compare  parameter estimation using three different
noise curves, advanced  LIGO, initial LIGO, 
and  VIRGO. 
The results of our study can be summarised as follows:
\begin{enumerate}
\item The parameter estimation of non-spinning binaries  improves significantly,
as expected, by employing the  3.5PN phasing formula instead of the 2PN one.
It is no surprise that the same trend is observed for all the three detectors.
Improvements are larger for NS-BH and BH-BH systems and least for NS-NS
binary. For initial LIGO, at a SNR of 10, the improvement in the estimation 
of parameters ${\cal M}$ and $\eta$ for BH-BH binaries is as large as 
19\% and 52\%, respectively, whereas for NS-BH binaries it is 15\% and 50\%. 
Improvements in the case of VIRGO are slightly less compared to 
LIGO (cf.\ Table~\ref{table:improvement}).
\item  In  proceeding from 1PN to 3.5PN, one sees an oscillation 
of variances with each half  PN order.  However, the errors in the 
mass parameters at 3.5PN are always smaller than at 1PN and one can see
a convergence within this limited sequence.  The oscillation 
of errors is a characteristic feature of the PN approximation. In
Ref.~\cite{AIRS}, a similar oscillatory behaviour is seen in the context of
the detection problem. The variation in parameter estimation accuracies with PN
orders seem to be dominantly determined by the covariances between the
parameters $t_c$, $\phi_c$, ${\cal M}$ and $\eta$.
\item For sources at a {\it fixed distance} the errors in the  
estimation of parameters are least for advanced LIGO and the
highest for initial  LIGO, the performance of VIRGO being in between. 
Although initial LIGO and VIRGO obtain similar SNRs for sources 
with the total mass in the range $[1,\, 50]M_\odot,$ the errors 
in VIRGO are smaller than in initial LIGO by a factor 2--4 
due entirely to its greater bandwidth of observation.
\item  
The  number of useful cycles is greater in VIRGO than LIGO
for higher mass binaries ($M\simeq10M_{\odot}$) but the opposite is true for lower
mass binaries. 
\item 
Parameter estimation is better if the number of
useful cycles is higher but the performance also depends on the
sensitivity bandwidth of the instrument.
The notion of number of useful cycles together with bandwidth 
can be used to gauge detector performance with regard to parameter estimation.
\item The variation of the  Fourier amplitude of the gravitational
waveform
across different PN orders arising from  its dependence on the 
frequency-sweep ${dF}/{dt}$, and its implication on 
parameter estimation is examined.
We present a Table showing how the  SNR varies across the  PN orders for 
the initial LIGO noise curve. 
This correction affects the errors associated with parameter
estimation by  less than 10\% and  motivates an analysis using the complete waveform
including  all other harmonic
contributions to the GW amplitude from the `plus' and
`cross' polarisations which are now available up to 2.5PN in  the comparable
mass case \cite{ABIQ1}.
\end{enumerate}
\subsection{Limitations, Caveats and Future directions}
We conclude by pointing out   the regime of validity of our
analysis  of error bounds, its limitations
and possible  future directions.
\begin{enumerate}
\item Our estimates are based on  the Cramer-Rao
bound which is valid only in the regime of high SNR. Though at a SNR of
10 our calculations may be reasonably secure, in general  they are less 
rigorous and provide only an upper bound on the errors involved. A full-fledged Monte-Carlo simulation would provide tighter bounds,
though that would be computationally quite expensive. 
\item In Sec~\ref{sec:non-RWF} we addressed the effect of inclusion of
amplitude corrections arising from the frequency-sweep. This treatment is
not fully consistent as one neglects the amplitude corrections from the
other  harmonics of the orbital frequency;  
a future study should address this issue
more consistently.
\item Based on the recent runs of the GW detectors LIGO and VIRGO more `realistic' noise curves are now
available. The parameter estimation using these
 realistic noise curves should be eventually addressed. 
\item A similar study in the case of spinning binaries is not possible until
the terms corresponding to the effect of  spins in the phasing formula are available beyond the present 2PN
accuracy. 
\item A more detailed study is needed for completely understanding of the reasons for
PN variations of the errors. We leave this for future study.
\item The higher order phasing terms could play a major role also in the
estimation of distance of the binary for a network of detectors. We
will  address this problem in a future work.
\end{enumerate}

While finalising this paper we learnt that E. Berti and A. Buonanno
have also looked at the estimation of parameters using the 3.5PN phasing
formula \cite{BB}.
\acknowledgments
 We thank the LIGO Scientific Collaboration for
 providing us an estimate of the Advanced LIGO noise curve.
We thank R.\ Balasubramanian, Luc Blanchet and Sanjeev Dhurandhar  for useful discussions, suggestions 
and critical comments on the manuscript. We are grateful to Alessandra
Buonanno for bringing to our notice an incorrect normalization used in
the log term at 3PN in the phasing formula in an earlier version of the
paper. K.G.A. would like to thank 
P.\ Ajith for  useful discussions.
P.A.S.\ thanks the Raman Research Institute for a VSR fellowship and hospitality.
B.R.I.\ thanks the University of Wales and Cardiff, Institut d'Astrophysique de Paris 
and Institut des Hautes \'Etudes Scientifiques, France  for hospitality during the 
final stages of the writing of the paper.  This research was supported partly by 
grants from the Leverhulme Trust and the Particle Physics and Astronomy Research 
Council, UK. BSS thanks the Raman Research Institute, India, for supporting his 
visit in July-August 2004 during which some of this research was carried out.
All the analytical as well as numerical calculations leading to the
results of the paper have been performed with the aid of {\it Mathematica}.


\begin{thebibliography}{99}
\bibitem{a1} http://www.ligo.caltech.edu/\\
 http://www.virgo.infn.it/\\http://www.geo600.uni-hannover.de/\\
http://tamago.mtk.nao.ac.jp/
\bibitem{Cutler-ThorneGR16}C.\ Cutler and K.S.\ Thorne,  Proceedings of
$16^{\rm th}$ international conference on General relativity and
Gravitation, Eds N.T.\ Bishop and S.D.\ Maharaj (World Scientific, 2002); Preprint {\tt gr-qc/0204090} (2002).
\bibitem{GWA} S.A.\ Hughes, Annals Phys. {\bf 303} 142 (2003); Preprint
{\tt astro-ph/0210481}; B.S.\ Sathyaprakash, {\tt gr-qc/0405136} (2004).
\bibitem{Luc-LivRev} L.\ Blanchet, Liv. Rev. Relativ. {\bf 5}, 3 (2002).
\bibitem{peters}C.\ Peters, Phys.\ Rev.\ {\bf136}, B1224 (1964).
\bibitem{Thorne} K.S.\ Thorne  in {\it 300 years of gravity}, edited by
 S.W.\ Hawking and W.\ Israel  (Cambridge University Press, Cambridge, England,1987) 330.
\bibitem{Helstrom}C.W.\ Helstrom, {\it Statistical Theory of Signal Detection}, (Pergamon, Oxford, England, 1968).  
\bibitem{Schutz} B.F.\ Schutz, In  {\it The Detection of Gravitational
Radiation}, Edited by D.\ Blair (Cambridge University Press, England, 1989).
\bibitem{phasing}L.\ Blanchet, T.\ Damour and B.R.\ Iyer, Phys.\ Rev.\
D{\bf 51}, 5360 (1995); {\bf 54}, 1860 (E) (1996); C.M.\ Will
and A.G.\ Wiseman, Phys.\ Rev.\ D {\bf 54}, 4813 (1996);
L.\ Blanchet, T.\ Damour, B.R.\ Iyer, C.M.\ Will and A.G.\ Wiseman, Phys.\ Rev.\ Lett. {\bf 74}, 3515 (1995);
L.\ Blanchet, Phys.\ Rev.\ D {\bf 54}, 1417 (1996);
L.\ Blanchet, Class.\ Quant.\ Grav.\ {\bf 15},  113 (1998);
L.\ Blanchet, B.R.\ Iyer and B.\ Joguet, Phys.\ Rev.\ D {\bf 65}, 064005
(2002).
\bibitem{BFIJ}L.\ Blanchet, G.\ Faye, B.R.\ Iyer  and B.\ Joguet,
Phys.\ Rev.\ D {\bf 65}, 061501(R) (2002);
\bibitem{BIWW}L.\ Blanchet, B.R.\ Iyer, C.M.\ Will and A.G.\ Wiseman, 
Class.\  Quant.\ Grav.\ {\bf 13}, 575 (1996)
\bibitem{ABIQ1} K.G.\ Arun, L.\ Blanchet, B.R.\ Iyer and M.S.S.\ Qusailah, 
Class.\  Quant.\ Grav.\ {\bf 21}, 3771 (2004); Preprint {\tt gr-qc/0404085}.
\bibitem {Sintes} A.M.\ Sintes  and A.\ Vecchio, gr-qc/0005058 (2000);
and {\tt gr-qc/0005059}(2000).
\bibitem {HM1}R.W.\ Hellings  and T.A.\ Moore, Class.\ Quant.\ Grav.\ {\bf 20}, S181 (2002)
\bibitem {HM2} T.A.\ Moore  and R.W.\ Hellings, Phys.\ Rev.\ D {\bf 65}, 062001 (2003).
\bibitem {CRB} C.R.\ Rao, Bullet. Calcutta Math. Soc. {\bf 37}, 81 (1945); 
H. Cramer, {\it Mathematical methods in statistics} (Princeton University Press, NJ, 1946)
\bibitem{KKT94} K.D. Kokkotas, A.\ Kr\'olak and G. Tsegas,
Class.\ Quant.\ Grav. {\bf 11}, 1901 (1994). 
\bibitem{BSS1}R.\ Balasubramanian, B.S.\ Sathyaprakash and S.V.\ Dhurandhar, Phys.\ Rev.\ D {\bf53}, 3033 (1996); {\bf 54}, 1860 (E) (1996) .
\bibitem{BSS2} R.\ Balasubramanian, B.S.\ Sathyaprakash and S.V.\ Dhurandhar, Pramana {\bf45}, L463 (1995)
\bibitem{BaDh98} R.\ Balasubramanian and  S.V.\ Dhurandhar, Phys.\ Rev.\ D {\bf 57}, 3408 (1998).
\bibitem{Nich-Vech} D.\ Nicholson and A.\ Vecchio, Phys.\ Rev.\ D {\bf 57},4588 (1998).
\bibitem{CF}  C.\ Cutler  and E.E.\ Flanagan, Phys.\ Rev.\ D {\bf 49}, 2658 (1994).
\bibitem{JK94} P. Jaranowski and A. Kr\'olak, Phys.\ Rev.\ D {49}, 1723
(1994).
\bibitem{JKKT96} P. Jaranowski, K.D. Kokkotas, A. Kr\'olak and G.
Tsegas, Class. \ Quant. \ Grav. {13}, 1279 (1996).
\bibitem{PW} E.\ Poisson and C.M.\ Will , Phys. Rev.D {\bf 52}, 848 (1995).
\bibitem{Krolak2}  A.\ Kr\'olak, K.D.\ Kokkotas  and G.\ Sch\"afer , Phys.\ Rev.\ D {\bf 52}, 2089 (1995).
\bibitem{BBW04} E.\ Berti, A.\ Buonanno and C.M.\ Will, {\tt gr-qc/0411129}
\bibitem{DJS01}T.\ Damour, P.\ Jaranowski and G.\ Sch\"afer, Phys. Lett. B
{\bf 513}, 147 (2001).
\bibitem{BDE04} L.\ Blanchet, T.\ Damour and G Esposito-Far\'ese,
Phys.\ Rev.\ D {\bf 69}, 124007 (2004).
\bibitem{BDEI04} L.\ Blanchet, T.\ Damour, G.\ Esposito-Far\`ese and  B.R.\ Iyer,
Phys.\ Rev.\ Lett. {\bf 93}, 091101 (2004).
\bibitem{BI04} L.\ Blanchet and B.R.\ Iyer,
Phys.\ Rev.\ D {\bf 71}, 024004 (2005); {\tt gr-qc/0409094}. 
\bibitem{BDI04} L.\ Blanchet, T.\ Damour and B.R.\ Iyer,
Class.\ Quant.\ Grav. {22}, 155 (2005); {\tt gr-qc/0410021}.
\bibitem{BDEI05} L.\ Blanchet, T.\ Damour, G.\ Esposito-Far\`ese and B.R.\ Iyer (2005); {\tt gr-qc/0503044}.
\bibitem{Itoh} Y.\ Itoh and T.\ Futamase, Phys.\ Rev.\ D
{\bf 68}, 121501(R) (2003); Y.\ Itoh, Phys.\ Rev.\ D {\bf 69}, 064018 (2004)
\bibitem{dis3}  T.\ Damour, B.R.\ Iyer and  B.S.\ Sathyaprakash, Phys.\ Rev.\ D {\bf 63}, 044023 (2001).
\bibitem{KWW} L. E. \ Kidder, Phys.\ Rev.\ D {\bf 52}, 821 (1995); L.E.\ Kidder, C.M.\ Will  and A.G.\ Wiseman, Phys.\ Rev.\ D {\bf 47}, R4183 (1993).
\bibitem{dis2} T.\ Damour, B.R.\ Iyer and  B.S.\ Sathyaprakash, Phys. Rev.D {\bf 62}, 084036 (2000).
\bibitem{SPA} K.S.\ Thorne  in \cite{Thorne}; S.V.\ Dhurandhar, A.
Kr\'olak , B.F.\ Schutz  and J.\ Watkins  (unpublished); B.S.\ Sathyaprakash,
Phys.\ Rev.\ D {\bf 50}, R7111 (1994);
S.\ Droz, D.J.\ Knapp, E.\ Poisson and B.J.\ Owen, Phys.\ Rev.\ D {\bf 59}, 124016 (1999).
\bibitem{Finn} L.S.\ Finn, Phys.\ Rev.\ D {\bf 46}, 5236 (1992).
\bibitem{Finn-Chernoff}  L.S.\ Finn and D.F.\ Chernoff, Phys.\ Rev.\ D {\bf 47}, 2198 (1993).
\bibitem{Luc-Sathya} L.\ Blanchet and B.S.\ Sathyaprakash, Class.\ Quant.\ Grav.\ {\bf 11}, 2807 (1994).
\bibitem{Wainstein} L.A.\ Wainstein  and  V.D.\ Zubakov, {\it Extraction of
Signals from Noise}, (Prentice-Hall, Englewood Cliffs, 1962).
\bibitem{Krolak1}A.\ Kr\'olak, J.A.\ Lobo and B.J.\ Meers, Phys.\ Rev.\ D {\bf 48}, 3451 (1993).
\bibitem{Davies}M.H.A.\ Davies, in {\it Gravitational Wave Data Analysis}, 
edited by B.F.\ Schutz (Kluwer Academic, Dordrecht, 1989). 
\bibitem{dis4} T.\ Damour, B.R.\ Iyer and  B.S.\ Sathyaprakash, Phys.\ Rev.\ D {\bf 66}, 027502 (2002).
\bibitem{Science} A.\  Abramovici {\it et.\ al.}, Science {\bf 256}, 325
(1992).
\bibitem{AIRS}  P.\ Ajith , B.R.\ Iyer, C.A.K.\ Robinson
and B.S.\ Sathyaprakash, Phys.\ Rev.\ D {\bf 71}, 044029 (2005).
\bibitem{BB}E. Berti and A. Buonanno (In preparation)
\end{thebibliography}
\end{document}